\documentclass[english,aps,pre,reprint,superscriptaddress, onecolumn, notitlepage, nofootinbib]{revtex4-1}
\usepackage[T1]{fontenc}
\usepackage[latin9]{inputenc}
\usepackage{babel}
\usepackage{amsmath, amsfonts, bm, framed}
\usepackage{comment}
\usepackage{amssymb}
\usepackage{graphicx}
\usepackage{float}
\usepackage{grffile}
\usepackage{algorithm}
\usepackage{algpseudocode}
\usepackage[dvipsnames]{xcolor}

\usepackage{ragged2e}
\makeatletter
\long\def\@makecaption#1#2{%
  \par
  \vskip\abovecaptionskip
  \begingroup
    \small\rmfamily
    \justifying
    \noindent
    \@make@capt@title{#1}{#2}\par
  \endgroup
  \vskip\belowcaptionskip
}
\makeatother

\makeatletter

\usepackage[unicode=true,pdfusetitle,
 bookmarks=true,bookmarksnumbered=false,bookmarksopen=false,
 breaklinks=false,pdfborder={0 0 1},backref=false,colorlinks=false]
 {hyperref}

\usepackage{braket}

\DeclareMathOperator{\sign}{sign}

\DeclareMathOperator*{\argmin}{argmin}

\usepackage{titletoc}

\titlecontents{section}
  [0em]                         
  {\addvspace{0.2em}}           
  {\contentslabel{1.1em}}       
  {\hspace*{2.5em}}             
  {\titlerule*[0.6pc]{.}\contentspage}

\titlecontents{subsection}
  [2.em]                       
  {}
  {\contentslabel{1.1em}}         
  {\hspace*{3em}}
  {\titlerule*[0.6pc]{.}\contentspage}

\makeatother

\begin{document}

\title{On the robustness of noisy solutions in non-convex neural networks}

\author{Enrico M. Malatesta}
\affiliation{Department of Computing Sciences, Bocconi University, Milano, Italy}
\affiliation{Bocconi Institute for Data Science and Analytics (BIDSA)}

\author{Alessandra Passalacqua}
\affiliation{Department of Computing Sciences, Bocconi University, Milano, Italy}

\author{Riccardo Zecchina}
\affiliation{Department of Computing Sciences, Bocconi University, Milano, Italy}
\affiliation{Bocconi Institute for Data Science and Analytics (BIDSA)}



\begin{abstract}
    Optimization in non-convex neural network models is strongly influenced by the geometry of the solution space: sparse, isolated, point-like clusters are typically algorithmically inaccessible, whereas wide and flat regions can be found efficiently despite being relatively rare. At zero temperature this picture has been formalized in binary perceptrons through the overlap gap property (OGP), which limits algorithmic access to configurations with zero training error above a critical constraint density $\alpha_{\rm OGP}$. Here we extend this description to finite temperature, where a positive training error is allowed and statistically penalized. We first show that the frozen one-step replica-symmetry-breaking solution, dominating the zero temperature equilibrium measure, survives at any finite temperature. We furthermore derive a general criterion, based on the smoothness of the single-pattern Gibbs weight near the decision boundary, that determines when a finite-temperature relaxation of the loss removes freezing. We then extend the OGP construction to finite temperature and show that dense, algorithmically accessible regions of finite-energy configurations persist  beyond $\alpha_{\rm OGP}$, up to a threshold $\alpha_{\rm OGP}(\epsilon)$ that grows with the allowed training error $\epsilon$. Finally, in the teacher-student setting, we show that these wide, finite-energy regions still retain good generalization. Using a finite energy message-passing algorithm, we  demonstrate numerically that thermal noise enables effective generalization in the regime of constraint densities where both recovering the teacher and finding a zero temperature solution are computationally hard.
\end{abstract}

\maketitle

\section{Introduction}
Over the last decade, numerous studies at the intersection of statistical physics and machine learning showed how the local geometry of the energy landscape of neural networks can be exploited to guide their optimization, and consequently improve their performance, despite the practical difficulty of navigating highly non-convex spaces \cite{subdominant, unreasonable, Entropy-SGD}. The robust ensemble formalism was introduced in the study of non-convex neural networks, to prove the existence of atypical, but algorithmically accessible and well-generalizing, wide and flat regions at zero training error, as opposed to inaccessible isolated, point-like solutions, and algorithms designed to search for these regions turned out to be successful. At the same time, mathematicians have introduced the rigorous theoretical framework of the overlap gap property (OGP): whenever the near-optimal solutions of a problem lie in small, or even point-like, disconnected clusters, these become unreachable to a broad class of optimization algorithms \cite{Gamarnik}.
The general formulation of the OGP theory makes it applicable to some well-known problems in statistical physics of complex systems \cite{Gamarnik-pspin}, random graph theory \cite{independent_set} and simple neural network models \cite{gamarnik2022sbp}.

This body of work, however, has almost exclusively been developed at zero temperature, where learning is phrased as a constraint satisfaction problem (CSP): a configuration has nonzero weight in the partition function only if it fits every training point correctly. A partial exception is offered by studies that relax the requirement by introducing a stability margin in the classification constraint, which enlarges the space of solutions of the CSP while remaining at zero temperature \cite{Baldassi_Typical_Atypical};
even so, both settings ultimately demand exact satisfaction of a fixed number of constraints. This is a demanding requirement,
at odds with how learning is typically carried out in practice, where a nonzero training error is common - and can even improve generalization and robustness to noise, in contrast to overfitting. A prominent example
is offered by large language models,
which are extremely heavy to train, so that
compute-optimal training generally stops before convergence,
and a nonzero training loss is the rule rather than the exception
\cite{kaplan2020, hoffmann2022}.


In this work we extend the study of the geometry of the configuration space to finite temperature, where imperfect classification is allowed but statistically penalized.
We consider the paradigmatic case of binary perceptrons, either storing random patterns or learning a classification rule from a teacher. The simplicity of the architecture allows a tractable analytical study, and yet the discreteness of the weights renders the solution space non-convex, endowing it with the rich geometrical structure that motivates our study.
Models are introduced in Section \ref{sec:models}, each being defined through its Hamiltonian.
For each model, we will consider two classes of optimal and near-optimal weights in the energy landscape, corresponding to equilibrium and out-of-equilibrium configurations.
The equilibrium ones are those typically encountered when sampling from the Gibbs measure with the standard error counting loss, at temperature $T$. At zero temperature, these are known to be organized in a frozen one-step replica-symmetry-breaking (1RSB) structure \cite{Krauth, aubin2019storage},  decomposed into exponentially many geometrically isolated, point-like clusters \cite{Kabashima, gamarnik2022sbp}.
The second class consists of atypical, out-of equilibrium configurations, obtained by biasing the measure to favor regions with large local entropy, that are invisible to the equilibrium analysis, but visible to common optimization algorithms because of their geometrical accessibility.

In Sec.~\ref{sec:dynamical_temperature_freezing} we start our discussion from the typical equilibrium configurations.
We ask whether raising the temperature is sufficient to dynamically unfreeze the landscape, and show that it is not: the frozen structure persists at every finite temperature, in agreement with the dynamical field theory analysis of \cite{Horner-storage}. We show that the dynamical temperature diverges in the thermodynamic limit. We identify the general mechanism responsible for freezing, related to the roughness of the Gibbs weight in the equilibrium measure: some modifications of the Hamiltonian can change the structure of the landscape, allowing the emergence of clusters even at the typical level.

Then, in Sec.~\ref{sec:ogp}, considering again the standard error-counting loss function, we move to the study of atypical dense regions of solutions. At zero temperature, these exist for constraint densities $\alpha$ smaller than a threshold $\alpha_{\rm OGP}$, and are algorithmically reachable (up to a small gap) \cite{unreasonable}. We show that the onset of the overlap gap property can be followed in temperature: dense flat regions still exist as thermal noise is injected,
but they exist up to a threshold $\alpha_{\rm OGP}(\epsilon)$ that grows with the allowed energy $\epsilon$.
Our approach is based on the replica method \cite{mezard1988spin} and it gives results that are compatible with the previous analytical estimates \cite{gamarnik2022sbp, stojnic2026parametric,stojnic2026ultrametric}.
We probe the accessibility of the near-optimal clusters numerically, with a message passing algorithm naturally devised to converge towards finite-energy flat regions in the energy landscape, even beyond the zero-temperature OGP threshold.
Finally, in Sec.~\ref{sec:following}, we focus on the teacher-student setting and test the generalization performance of the finite-energy wide clusters, showing that the thermal noise in training error does not obstruct learning.
On the contrary: beyond the zero-temperature OGP threshold, restricting the search to exact-fit configurations can create an algorithmic obstruction, whereas accessible finite-error dense regions may still generalize well.
This suggests that it is beneficial to shape optimization algorithms to search for wide regions rather than isolated global minima, even when the former carry finite energy.


\section{Models and finite-temperature ensembles}
\label{sec:models}

We consider a class of binary perceptron models with $N$ Ising weights $\bm w\in\{-1,+1\}^N$, trained on datasets comprising $P=\alpha N$ random patterns divided into two classes. The input vectors have independent standard Gaussian entries,
$x_i^\mu\sim\mathcal N(0,1)$, and we denote by
\begin{equation}\label{main:def_field}
    h^\mu(\bm w)
    =
    \frac{1}{\sqrt N}\sum_{i=1}^N w_i x_i^\mu
\end{equation}
the local field associated with the pattern $\mu$. The quantity
\begin{equation}
    s^\mu(\bm w)
    =
    y^\mu h^\mu(\bm w),
\end{equation}
is the stability of the input $\boldsymbol{x}^\mu$,
where $y^\mu = \pm 1$ is the binary label associated to the input pattern.
In the case of the asymmetric binary perceptron (ABP), the perceptron is said to classify the pattern $\boldsymbol{x}^\mu$ correctly whenever $s^\mu(\bm w)>0$. We will call $\boldsymbol{w}$ a \emph{solution} of the ABP problem if $s^{\mu}(\boldsymbol{w}) > 0$ for any $\mu \in [P]$.

In the following we shall consider two classical ways of generating the labels.  In the \emph{storage} setting the labels $y^\mu = \pm 1$ are independent Rademacher random variables. Since the distribution of the patterns is symmetric, the labels can then be gauged away and one may set $y^\mu=1$ for all $\mu$ without loss of generality. In this setting, one can focus on the \emph{optimization} task, by which we mean finding student configurations $\boldsymbol{w}$ that achieve zero training error, at constraint density $\alpha$.
In the \emph{teacher-student} problem, instead, the labels are generated by a planted binary teacher $\bm w^\star\in\{-1,+1\}^N$ via
\begin{equation}
    y^\mu
    =
    \sign\left(
    \frac{1}{\sqrt N}\sum_{i=1}^N w_i^\star x_i^\mu
    \right).
\end{equation}
In the teacher-student setting, our main focus will be again the optimization task.
This has to be distinguished from the \emph{inference} task, where one is interested in inferring the planted signal~\cite{Gyorgyi1990,Gardner_1989}. The teacher-student setting also allows to study the \textit{generalization} capability of the network.
In particular, the teacher-student overlap $r(\bm w,\bm w^\star) = \frac{1}{N}\sum_{i=1}^N w_i w_i^\star$ determines the generalization error of the student, which is the probability that the student and the teacher disagree on a previously unseen Gaussian input, through
\begin{equation}
    \epsilon_g(r)
    =
    \frac{1}{\pi}\arccos r\,.
\end{equation}
Both the storage and teacher-student settings have been studied extensively in the statistical mechanics literature, see~\cite{GardnerDerrida, Gardner_1989, Krauth, Gyorgyi1990, Opper1991, Kabashima, subdominant,unveiling2021,Baldassi_Typical_Atypical}.

While our main focus is the optimization problem, it is also useful to consider configurations with a finite training error. To this end, we introduce a Gibbs measure over the weight configurations, where each pattern contributes through a single-pattern Boltzmann factor $\mathcal K$. The Gibbs partition function is
\begin{equation}
    Z_{\mathcal D}(\beta)
    =
    \sum_{\bm w\in\{\pm1\}^N}
    \prod_{\mu=1}^{\alpha N}
    \mathcal K\left(s^\mu(\bm w)\right),
    \label{eq:generic_partition_main}
\end{equation}
where the dependence on the inverse temperature ($\beta$) is encoded in $\mathcal K$. In the zero-temperature limit, suitable choices of $\mathcal K$ recover the hard-constraint optimization problem discussed above. The quenched free entropy density associated to the partition function is:
\begin{equation}
    \label{eq::free_entropy}
    \phi(\beta)
    =
    \lim_{N\to\infty}\frac1N
    \mathbb E_{\mathcal D}\log Z_{\mathcal D}(\beta).
\end{equation}
The model we will focus on in this paper is the ABP, with single Gibbs weight given by
\begin{equation}
    \mathcal K^{\rm ABP}(s) = e^{-\beta\Theta(-s)} = e^{-\beta}+ \left(1-e^{-\beta}\right)\Theta(s).
    \label{eq:K_ABP_main}
\end{equation}
In the zero-temperature limit, the measure is uniform on zero-error solutions whenever such solutions exist, while at positive temperature it also assigns weight $e^{-\beta}$ to each violated constraint. Equivalently, this corresponds to weighting configurations $\boldsymbol{w}$ according to a Gibbs distribution having as energy a loss function $\mathcal L_{\mathcal D}$
\begin{equation}
    Z_{\mathcal D}(\beta) = \sum_{\bm w} e^{-\beta \mathcal L_{\mathcal D}(\bm w)}, \qquad \mathcal L_{\mathcal D}(\bm w) = \sum_{\mu=1}^{\alpha N}
    \Theta\left(-s^\mu(\bm w)\right).
    \label{eq::uncloned_partition_function}
\end{equation}
which counts the number of misclassified training patterns. The same formalism is generic and can be used also to study other models. For example, similar results that we will present for the ABP will be valid for the symmetric binary perceptron (SBP)~\cite{aubin2019storage, gamarnik2022sbp} as well and will be discussed in the appendices.  In the SBP with margin $\kappa>0$ the labels play no role and the constraint requires the local field to lie inside a window of width $2\kappa$:
\begin{equation}
    |s^\mu(\bm w)| = |h^\mu(\bm w)|\le \kappa.
\end{equation}
The single-pattern Gibbs weight corresponding to an error counting loss function is therefore
\begin{equation}
    \mathcal K^{\rm SBP}(s)
    =
    e^{-\beta\Theta(|s|-\kappa)}
    =
    e^{-\beta}+\left(1-e^{-\beta}\right)\Theta(\kappa-|s|).
    \label{eq:K_SBP_main}
\end{equation}
Other choices of $\mathcal K$ may impose the same hard constraint in the limit $\beta\to\infty$, but they can lead to different finite-temperature physics.  This point is central for the discussion of the
next section.  

The free entropy in equation~\eqref{eq::free_entropy} can be computed using the replica method~\cite{mezard1988spin}. The derivations using a Replica-Symmetric (RS) or a 1-step Replica Symmetry Breaking (1RSB) Ansatz are standard in the statistical physics literature and the detailed derivations are reported for convenience of the reader in Appendix~\ref{sec::free_entropy}.

\section{Dynamical temperature and freezing}
\label{sec:dynamical_temperature_freezing}

At zero temperature the equilibrium measure of the ABP and SBP, equipped with the Gibbs weight \eqref{eq:K_ABP_main} and~\eqref{eq:K_SBP_main} respectively, is known to be described by a frozen one-step replica-symmetry-breaking (1RSB) structure~\cite{Kabashima,aubin2019storage,perkins2024,Barbier_2024,Barbier2025}.
In this picture the Gibbs measure decomposes into exponentially many clusters which are separated by extensive Hamming distances, while each individual cluster is point-like.
In replica notation this means
that the intra-state overlap between pairs of solutions is
\begin{equation*}
    q_1=1,
\end{equation*}
whereas the inter-state overlap $q_0$ remains strictly smaller than
$q_1$. In the SBP, the symmetry $\bm w\mapsto -\bm w$ further implies $q_0=0$, i.e. the point-like clusters are orthogonal.

A natural question is whether this frozen structure is exclusively a zero-temperature property, or whether it persists when positive training error configurations are assigned finite Gibbs weight. More generally, we ask how this equilibrium picture depends on the specific choice of the finite-temperature single-pattern Gibbs weight $\mathcal{K}$. We can answer those questions by computing the so-called \emph{dynamical temperature} $T_d$. This is defined as the largest temperature below which the 1RSB saddle-point equations admit a non-trivial solution with the Parisi block parameter $m=1$ and $q_1>q_0$ \cite{KirkpatrickThirumalai1987}. If, in addition, one has $q_1=1$, then for $T<T_d$ the corresponding 1RSB phase is frozen.

For the error-counting single-pattern weights $\mathcal K_{\rm ABP}$ and $\mathcal K_{\rm SBP}$, the computation reported in Appendix~\ref{app:dynamical-temperature} shows that, for every fixed $\alpha>0$ and every inverse temperature $\beta>0$, the $m=1$ 1RSB equations admit a non-trivial frozen solution. Therefore, the dynamical temperature diverges in the thermodynamic limit. In other words, the equilibrium measure is dynamically glassy at every finite temperature. This result is in agreement with what Horner found for the ABP in the storage setting using dynamical field theory techniques~\cite{Horner-storage}.

\begin{figure}[t]
    \centering
    \includegraphics[width=0.45\linewidth]{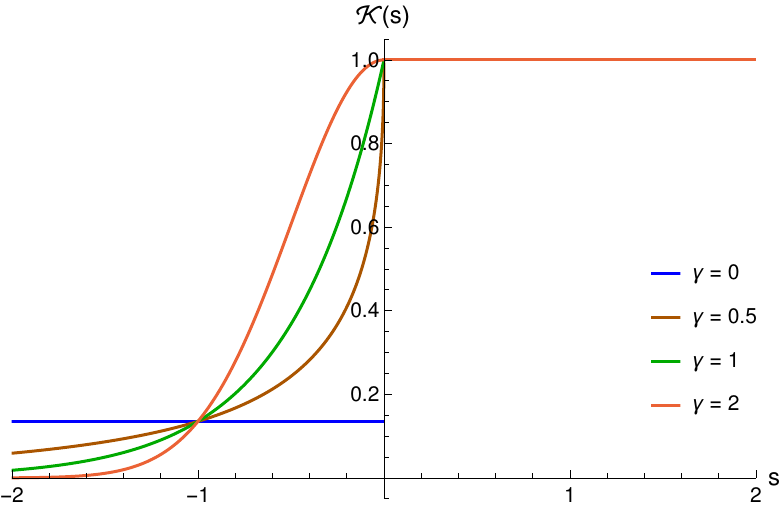}
    \hspace{1cm}
    \includegraphics[width=0.45\linewidth]{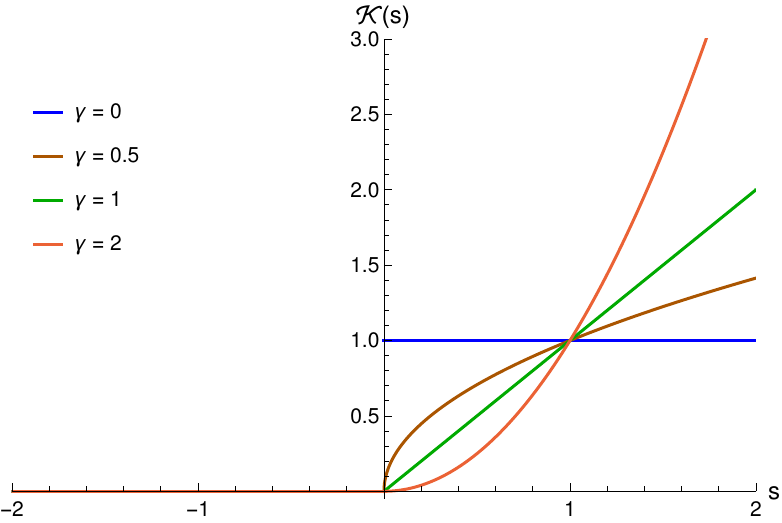}
    \caption{Plot of the single pattern Gibbs weight. Left: $\mathcal K(s) = e^{-\beta(-s)^\gamma \Theta(-s)}$ for $\beta = 2$ and $\gamma = 0$, 0.5, 1, 2. This form of the single pattern weight has been in considered by Horner~\cite{Horner-storage} for integer values of $\gamma$. Right: $\mathcal K(s) = s^\gamma\Theta(s)$ considered in~\cite{straziota2026generative} for the same values of $\gamma$ as in the left panel. In both the left and right plot, $\mathcal{K}$ has a jump discontinuity at the decision boundary $b=0$ for $\gamma = 0$, which induces freezing. In the left plot the discontinuity is recovered also in the large $\beta$ limit.  }
    \label{fig:single_pattern_weights}
\end{figure}

The mechanism leading to this result is quite general, and clarifies which features of the finite-temperature Gibbs weight are responsible for freezing. As detailed in Appendix~\ref{app::generic_criterion_freezing},
the 1RSB saddle point equations involve the function
\begin{equation}
    \mathcal H(x,y)
    =
    \int Dz\,\mathcal K(x+yz)\,,
    \label{eq:H_kernel_freezing}
\end{equation}
where $Dz=\frac{dz}{\sqrt{2\pi}}e^{-z^2/2}$ and $y=\sqrt{1-q_1}$.
The limit $q_1\to1$ therefore corresponds to the limit $y\to0$. Possible singularities in this limit can only come from the decision boundaries, namely from the points where the zero-temperature constraint changes value and where $\mathcal K$ may be non-smooth. We denote such a boundary by $b$. For example, in the ABP one has $b=0$, while in the SBP with margin $\kappa$ the boundaries are $b=\pm\kappa$. A crucial quantity controlling the onset of freezing is the boundary layer term
\begin{equation}
    \mathcal B_{\mathcal K}(b,y) \equiv
    y\int du\,
    \frac{
        \left[\partial_x\mathcal H(b+yu,y)\right]^2
    }{
        \mathcal H(b+yu,y)
    } .
    \label{eq:boundary_layer_integral_main}
\end{equation}
In particular, if
\begin{equation}
    \mathcal B_{\mathcal K}(b,y)\to\infty
    \qquad
    \text{as}
    \qquad
    y\to0,
\end{equation}
we argue in Appendix~\ref{app::generic_criterion_freezing} that the solution space is frozen. Conversely, if
$\mathcal B_{\mathcal K}(b,y)$ remains finite, or vanishes, as $y\to0$, then the frozen phase is absent. In that case a non-trivial 1RSB solution with $q_1<1$ may still appear, which does not
describe point-like clusters. We illustrate below such criterion on three types of single-pattern weights $\mathcal{K}$ shown in Fig.~\ref{fig:single_pattern_weights} that have been considered in the literature.

\begin{enumerate}
    \item \textbf{Jump discontinuity.} Consider a single pattern Gibbs weight that has a jump discontinuity at the decision boundary. $\mathcal K^{\rm ABP}$ and $\mathcal K^{\rm SBP}$ given in~\eqref{eq:K_ABP_main} and~\eqref{eq:K_SBP_main} fall in this class: for any $\beta>0$, they are discontinuous at the decision boundary, namely at the origin for the ABP and at $\pm \kappa$ for the SBP.  As we show in~\ref{app::generic_criterion_freezing}, this is responsible for the divergence of $\mathcal{B}_\mathcal{K}$ when $y\to 0$
          \begin{equation}
              \mathcal B_{\mathcal K}(b,y)
              \simeq
              \frac{C(\beta)}{y} \,,
          \end{equation}
          where $C(\beta)$ is a positive constant for every $\beta>0$, and
          vanishes only at $\beta=0$. This is the reason why the frozen solution dominates the Gibbs measure at all finite temperatures.

    \item \textbf{Horner's weight.}
          A different finite-temperature continuation has been considered first by Horner~\cite{Horner-storage} and more recently in~\cite{copycat} in the case of the ABP. This is obtained by replacing the error-counting loss by a soft penalty for violated constraints
          \begin{equation}
              \label{eq::HornerABP}
              \mathcal K(s)= e^{-\beta(-s)^\gamma \Theta(-s)} \,;
          \end{equation}
          see the left panel of Figure~\ref{fig:single_pattern_weights} for a plot. We are considering here for simplicity the case of the ABP, but the considerations we are making can be extended to any model. Note that for $\gamma=0$ equation~\eqref{eq::HornerABP} reduces to the error-counting weight in~\eqref{eq:K_ABP_main}, so we should expect freezing. For $\gamma>0$ and $\beta < \infty$, the Gibbs weight~\eqref{eq::HornerABP} is instead continuous at the decision boundary. In this case the boundary layer term scales as
          \begin{equation}
              \mathcal B_{\mathcal K}(b,y)
              \sim
              y^{2\gamma-1}.
          \end{equation}
          when $y \to 0$.
          Hence freezing is expected when $\gamma<1/2$, while for $\gamma>1/2$ the boundary layer is not singular enough to produce a frozen solution. A dynamical transition to a non-frozen phase may still be present at a finite temperature (and indeed it occurs, see~\cite{Horner-storage}). Finally, since in the limit $\beta \to \infty$ one has $ \mathcal K(s) \to \Theta(s)$, the frozen solution is restored at zero temperature.

    \item \textbf{Logarithmic potential.}
          Another class consists of Gibbs weights that vanish as a power law when the decision boundary is approached from within the feasible region. For example, consider:
          \begin{equation}
              \mathcal K(s) = s^\gamma \Theta(s)
          \end{equation}
          This should be distinguished from the soft-penalty form in Eq.~\eqref{eq::HornerABP}, which smooths the cost of violated constraints. Here, instead, the weight suppresses configurations that satisfy the constraint only marginally, assigning larger weight to configurations deeper inside the feasible region.

          Equivalently, this choice induces, within the feasible region, a logarithmic potential $V(s) = -\log \mathcal K(s) = -\gamma \log s$, which penalizes small positive stabilities and favors configurations farther from the decision boundary~\cite{straziota2026generative}. In this case one can show that the boundary layer term $ \mathcal B_{\mathcal K}$ behaves as
          \begin{equation}
              \mathcal B_{\mathcal K}(b,y)
              \sim
              y^{\gamma-1}.
          \end{equation}
          Therefore the frozen solution is expected for $\gamma<1$, while it
          is absent for $\gamma>1$. This picture has also been observed in~\cite{straziota2026generative} using a Franz-Parisi potential approach.
\end{enumerate}

\section{The Overlap Gap Property}
\label{sec:ogp}

The frozen equilibrium picture described in the previous section does not by itself imply that the problem of finding configurations $\boldsymbol{w}$ with a given energy level is algorithmically hard. The reason is that efficient algorithms do not need to sample from the equilibrium Gibbs measure. They may instead operate out of equilibrium, following atypical trajectories in configuration space and exploiting regions that are exponentially subdominant in the equilibrium measure. This distinction is particularly important in the case of the binary perceptron with error counting loss~\eqref{eq:K_ABP_main} which, as pointed out in the previous section, has a frozen landscape at any finite temperature and positive constraint density.

In the past decade, the statistical-physics literature has shown that message-passing algorithms, especially when combined with reinforcement or with local-entropy biases, can efficiently find solutions and improve the algorithmic thresholds of many different constraint satisfaction problems, ranging from the $K$-SAT and perceptron model~\cite{Braunstein_2006,Baldassi_2015,barbier2025finding}, to the graph coloring problem~\cite{Angelini2023,Budzynski_2019,Budzynski_2020}. In~\cite{subdominant} it has been conjectured that algorithmic success is associated with the existence of dense regions of solutions which, despite being rare and invisible to the equilibrium measure, are basins of attraction much larger than typical equilibrium states.
A recent theoretical explanation for this effectiveness was that imposing local entropy biases could delay the dynamical transition~\cite{Budzynski_2019,Budzynski_2020}, with random hypergraph bicoloring constituting a notable exception~\cite{Angelini2025}.

This physical picture has recently been reformulated, in the mathematical literature, through the framework of the overlap gap property (OGP). OGP is a geometric obstruction in the space of near-optimal configurations. Such a topological obstruction has been used to explain algorithmic barriers in random optimization problems \cite{independent_set} and, more recently, in perceptron-type models~\cite{Gamarnik,gamarnik2022sbp,benedetti25}. Informally, $m$-OGP holds when relevant $m$-tuples of configurations, organize into separated regions: they can be mutually close or mutually far, but there is an entire interval of intermediate overlaps in which no such configurations can be found. This gap provides an obstruction to broad classes of so-called \emph{stable} algorithms,
namely algorithms whose outputs remain close, with high probability and under a common random seed, when applied to sufficiently close or correlated instances of the random problem.

Here we study the OGP by considering $m$ real replicas, or \emph{clones}, of the system constrained to have fixed mutual overlap. We limit ourselves here for simplicity to the ABP with the error counting loss~\eqref{eq:K_ABP_main} but the technical derivations presented in Appendix~\ref{sec::1RSB_entropy} are general. For $m\ge 2$ and $q_1\in[-1,1]$ the cloned partition function is defined as
\begin{equation}
    \begin{split}
        Z_{m,\mathcal D}(q_1;\beta)
        = \sum_{\{\bm w^a\}_{a=1}^{m}} \prod_{a=1}^{m}
        e^{-\beta \mathcal L_{\mathcal D}(\bm w^a)}
        \prod_{1\le a<b\le m} \delta\left( q_1-\frac{1}{N}\sum_{i=1}^{N}w_i^a w_i^b \right).
    \end{split}
    \label{eq::cloned_partition}
\end{equation}
The corresponding quenched free entropy per clone is
\begin{equation}
    \label{eq::free_entropy_clones}
    \phi_m(q_1;\beta)
    =
    \lim_{N\to\infty}
    \frac{1}{Nm}
    \mathbb E_{\mathcal D}
    \ln Z_{m,\mathcal D}(q_1;\beta).
\end{equation}
This quantity can be computed with the replica method. At the RS level, the result can be found simply by imposing a 1RSB structure on the equilibrium measure~\eqref{eq::uncloned_partition_function} and treating both $m$ and $q_1$ as external parameters rather than variational ones~\cite{Monasson1995,shaping}, see Appendix~\ref{sec::1RSB_entropy}. At zero temperature, $Z_{m,\mathcal D}(q_1;\infty)$ counts $m$-tuples of
solutions at mutual overlap $q_1$.
Thus, if for $\alpha$ larger than a certain threshold $\alpha_m(q_1)$ one has
\begin{equation}
    \label{eq::forbidden_overlap}
    \phi_m(q_1;\infty)<0,
\end{equation}
then with high probability no such $m$-tuple exists.
The onset of the $m$-OGP threshold is therefore obtained by finding the minimal value of the constraint density $\alpha$ for which a forbidden interval of overlaps~\eqref{eq::forbidden_overlap} starts being non-empty:
\begin{equation}\label{eq::minq1}
    \alpha_{\rm OGP}(m) = \min_{q_1} \alpha_m(q_1).
\end{equation}
Denote with $q_1^{\rm OGP}(m)$ the overlap that achieves the minimization.
Now, making the $\alpha$-dependence explicit, $\phi_m = \phi_m(q_1, \alpha; \beta)$, and assuming differentiability of the curve $\alpha_m(q_1)$, the free entropy vanishes on $\alpha =\alpha_m(q_1)$: $\phi_m(q_1, \alpha_m(q_1); \infty) = 0$ for all values of $q_1$. Differentiation gives:
\begin{align*}
    0 = \partial_{q_1}\phi_m + \alpha_m^\prime \partial_\alpha \phi_m \,,
\end{align*}
where the derivatives of $\phi_m$ are computed in $(q_1, \alpha_m(q_1))$.
Then, at $q_1 = q_1^{\rm OGP}(m)$, where $\alpha_m^\prime = 0$, we have $$\partial_{q_1}\phi_m(q_1;\infty) = 0\,.$$
To summarize, the two following conditions are sufficient to determine the values of $q_1^{\rm OGP}(m)$ and of $\alpha_{\rm OGP}(m)$, and therefore the OGP threshold:
\begin{subequations}
    \label{eq:ogp_zeroT_conditions}
    \begin{align}
        \phi_m(q_1;\infty)               & = 0,   \\
        \partial_{q_1}\phi_m(q_1;\infty) & = 0\,.
    \end{align}
\end{subequations}

At finite temperature the cloned partition function no longer counts configurations of zero training error. Rather, it weights configurations proportionally to their Boltzmann weight, at inverse temperature $\beta$. The entropy $s_m(q_1; \beta)$, i.e. the log-number of configurations that dominate the measure, can be found by a Legendre transform of the free entropy:
\begin{equation}
    s_m(q_1;\beta) = \phi_m(q_1;\beta) + \beta \alpha \epsilon_m(q_1; \beta) \,,
\end{equation}
where
\begin{equation}
    \epsilon_m(q_1; \beta) = -\frac{1}{\alpha} \frac{\partial \phi_m(q_1;\beta)}{\partial \beta} = \mathbb{E}_{\mathcal{D}} \left \langle \frac{1}{\alpha N m} \sum_{a=1}^m \mathcal{L}_{\mathcal{D}}(\bm{w}^a) \right\rangle_{\mathcal{D}}
\end{equation}
represents the average training error per pattern of the $m$ clones sampled from the constrained Gibbs measure 
induced by~\eqref{eq::cloned_partition}.
The finite-temperature OGP threshold is therefore defined by the solution of the system of equations analogous to \eqref{eq:ogp_zeroT_conditions}:
\begin{subequations}
    \label{eq:ogp_finiteE_conditions}
    \begin{align}
        s_m(q_1;\beta)               & = 0,   \\
        \partial_{q_1}s_m(q_1;\beta) & = 0\,.
    \end{align}
\end{subequations}

\begin{figure}[t]
    \centering
    \includegraphics[width=0.5\linewidth]{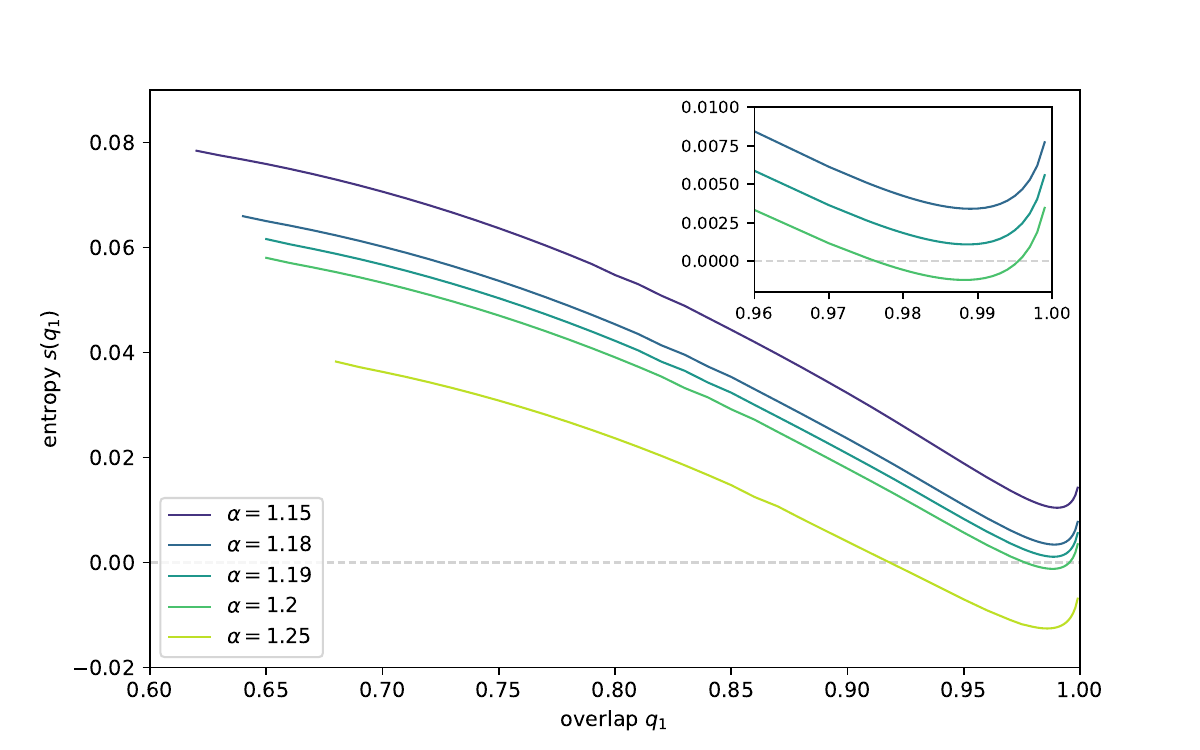}\hfill
    \includegraphics[width=0.5\linewidth]{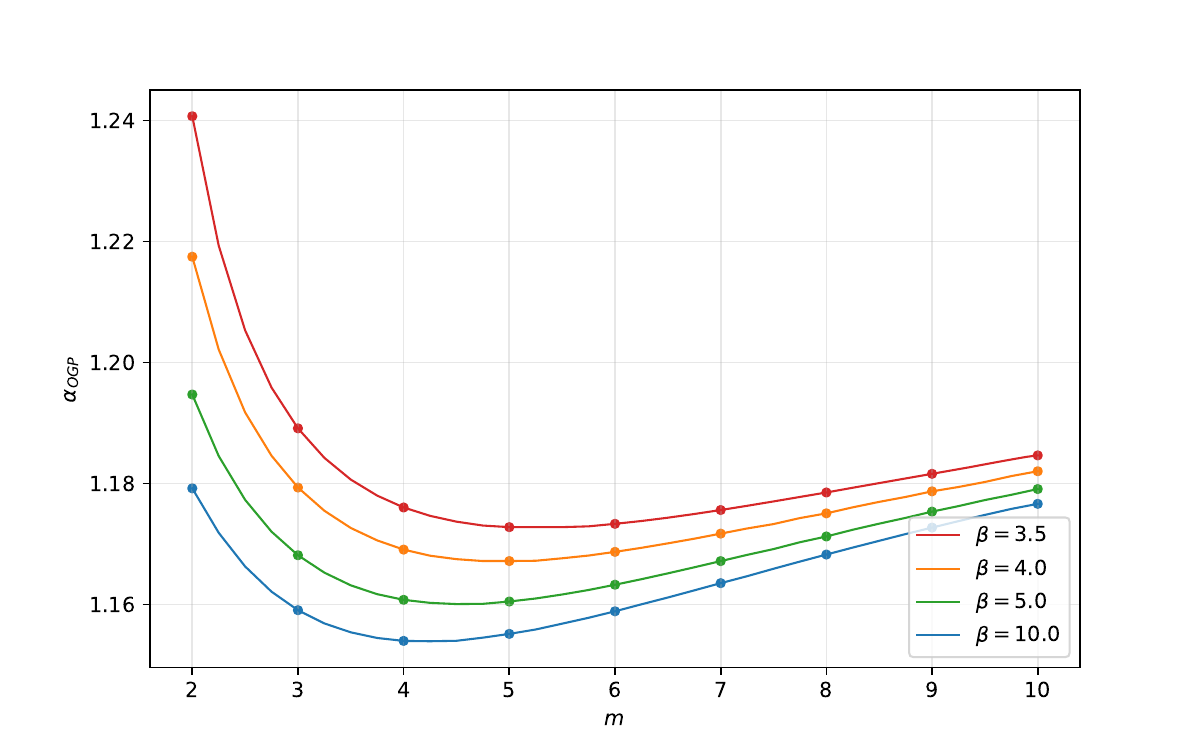}
    \caption{Emergence of the OGP at finite temperature in the ABP, teacher-student setting. The left panel depicts the entropy $s_m(q_1; \beta)$ for $m=2$ and $\beta = 5.0$, versus the overlap between the two clones $q_1$. For these values of $m$ and $\beta$, the OGP threshold is $\alpha_{\rm OGP}\simeq 1.195$ and $q_1^{\rm OGP}(m) \simeq 0.988$. Notice how, for $\alpha > \alpha_{\rm OGP}$, there is an entire interval of $q_1$'s for which the entropy is negative, meaning that, with high probability at large $N$, pairs of configurations satisfying the energy constraint exists at either small or very large overlaps, with an entirely forbidden interval of intermediate Hamming distances. The inset shows a zoom of the large  $q_1$ region. Right: $\alpha_{\rm OGP}$ vs $m$, for different values of $\beta$. The change in monotonicity that occurs at large $m$ is non-physical, and conjectured to be due to the RS Ansatz for the computation of the quenched free entropy; however, at higher temperatures the change of slope becomes progressively weaker.}
    \label{fig:find_OGP}
\end{figure}

The left panel of Figure \ref{fig:find_OGP} shows the entropy $s_m(q_1; \beta)$ for $m=2$ and a positive temperature, as a function of the overlap between pairs of solutions $q_1$, for different values of $\alpha$. The entropy is a non-monotonic function of $q_1$. As $\alpha$ is increased, the entropy curves shift downward. 
The OGP threshold is identified as the smallest value of $\alpha$ at which a forbidden interval of overlaps appears, namely when the minimum of the entropy curve as a function of $q_1$ first touches zero.

The right panel of Figure \ref{fig:find_OGP} shows $\alpha_{\rm OGP}$ as a function of $m$. For all the values of $\beta$ we have plotted, $\alpha_{\rm OGP}(m)$ is a non-monotonic function of $m$.
\begin{figure}[t]
    \centering
    \includegraphics[width=0.6\linewidth]{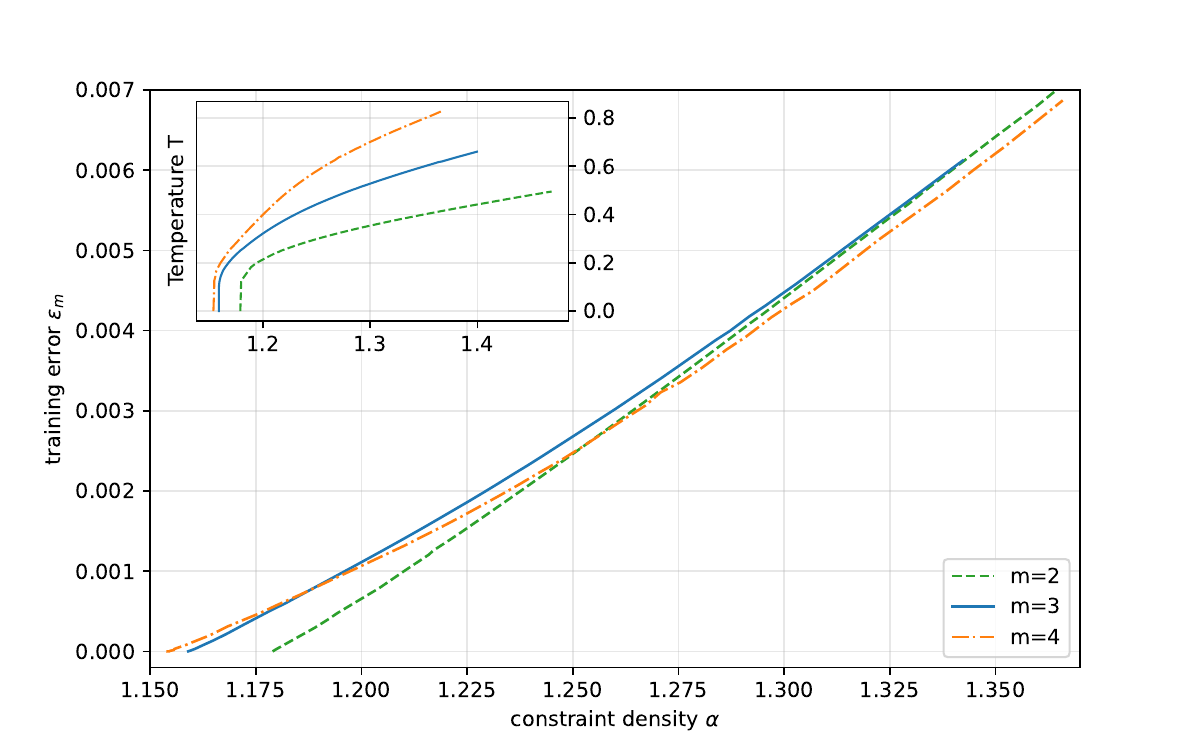}
    \caption{ OGP thresholds in the training error vs $\alpha$ plane. For each training error, the Overlap Gap Property holds for $\alpha>\min_m\alpha_{\rm OGP}(m)$. The inset plot shows the same OGP thresholds but in the $\alpha - T$ plane. }
    \label{fig:phase_diagram}
\end{figure}
The OGP threshold $\alpha_{\rm OGP}(m)$, is however expected to be a non-increasing function of $m$. Indeed, if for a given value of $\alpha$ there exist $m$ configurations $\bm w^1,\ldots,\bm w^m$ satisfying the prescribed single-replica energy constraint and the required pairwise overlap constraints, then any subset of $m'<m$ among them automatically satisfies the same conditions. Hence the existence of an admissible $m$-tuple implies the existence of an admissible $m'$-tuple if $m'<m$. Consequently, increasing $m$ can only make the constrained replicated problem in~\eqref{eq::cloned_partition} harder to satisfy, and the OGP threshold is expected to be non-increasing:
\begin{equation}
    \alpha_{\rm OGP}(m)
    \leq
    \alpha_{\rm OGP}(m')
    \qquad \text{for}\quad m'<m \,.
\end{equation}
The non-monotonic dependence on $m$ observed in the RS computation is therefore non-physical, and should be interpreted as an artifact of the RS Ansatz for the replicated constrained problem. The same phenomenon was already observed at zero temperature for both the ABP, SBP and in related models~\cite{gamarnik2022sbp,benedetti25, benedetti2026}. Resolving this unphysical behavior requires a detailed analysis of the replica-symmetry-breaking effects in the measure of the $m$-cloned system; this analysis will be presented elsewhere~\cite{Gianmarco2027}. At finite temperature the artifact persists, but it becomes progressively less pronounced. More precisely, let
\begin{equation}
    m_\star = \argmin_m \alpha_{\rm OGP}(m)
\end{equation}
denote the value of $m$ beyond which the RS estimate of the OGP threshold starts increasing. We observe that $m_\star$ shifts to larger values as the temperature is increased, indicating that the onset of the non-physical non-monotone regime is delayed. Some additional figures for the SBP showing a similar phenomenology are reported in Appendix~\ref{sec::1RSB_entropy}. Notice that the corresponding constraint density threshold
\begin{equation}
    \alpha_{\rm OGP} = \min_m \alpha_{\rm OGP}(m)
\end{equation}
is an upper bound to the appearance of an overlap gapped phase. Determining the exact value of the OGP threshold requires a more refined replica Ansatz capable of eliminating the non-physical non-monotonic dependence of $\alpha_{\rm OGP}(m)$ on $m$.

Figure~\ref{fig:phase_diagram} summarizes the finite-temperature OGP threshold in the ABP, in the teacher-student setting. The plot shows, as a function of the constraint density $\alpha$, the training error $\epsilon_m$ at which the OGP first appears for several values of $m$. Equivalently, the curve separates, for any value of $\alpha$, a large training error region, which is expected to be algorithmically accessible, from a low training error region, where we expect algorithmic hardness because of the presence of an overlap gap. In the right panel of Figure~\ref{fig:algorithmic_phase_diagram} we show similar OGP curves in the storage case. In the limit $\epsilon\to0$ (or correspondingly $\beta \to \infty$), we find in the teacher-student setting $\alpha_{\rm OGP} \simeq 1.154 
$, that is the zero-temperature threshold previously found in~\cite{subdominant}.
In the storage setting we similarly find $\alpha_{\rm OGP} \simeq 0.784$ 
~\cite{benedetti25} which is compatible with the results found in~\cite{subdominant,unveiling2021} using different approaches.

\subsection{Numerical simulations}

We now compare the finite-temperature OGP thresholds with the performance of an explicit algorithmic search for low-energy configurations. 
For a given value of the constraint density $\alpha=P/N$, we look for binary configurations with empirical training error
\begin{equation}
    \epsilon(\bm w)
    =
    \frac{1}{P}\sum_{\mu=1}^P
    \Theta\left(
    - \frac{y^\mu}{\sqrt N}  \boldsymbol{w} \cdot \boldsymbol{x}^\mu 
    \right)
\end{equation}
smaller than a prescribed threshold $\epsilon^{\rm targ}$.
Our starting point is Approximate Message Passing (AMP) applied to the finite-temperature Gibbs measure. At fixed inverse temperature $\beta$, AMP gives an estimate to the one-site marginal magnetizations
\begin{equation}
    a_i = \langle w_i\rangle_\beta \,,
\end{equation}
where $\langle \bullet \rangle_\beta$ denotes the average over the Gibbs measure at finite temperature. Equivalently, it computes the local fields $h_i$ such that $a_i=\tanh h_i$. A binary candidate configuration is then obtained by
\begin{equation}
    \label{eq::binary_estimator}
    \boldsymbol{w}_\star=\sign(\boldsymbol{a}).
\end{equation}
However, the standard AMP algorithm does not necessarily produce strongly polarized magnetizations $a_i$. This implies that the binary estimator in equation~\eqref{eq::binary_estimator} may not correspond to a low-error configuration.

\begin{algorithm}[H]
    \caption{Fixed-temperature reinforced AMP}
    \label{alg:ramp_fixed_beta}
    \begin{algorithmic}

        \Statex \textbf{Input:} patterns $\{\bm x^\mu,y^\mu\}_{\mu=1}^P$,
        inverse temperature $\beta$, target training error $\epsilon^{\rm targ}$,
        reinforcement parameter $\rho$, maximum number of iterations $t_{\max}$.

        \Statex \textbf{Initialize:} AMP local fields $\bm h^{t=0} \in \mathbb{R}^N$, marginal magnetization $\bm a^{t=0}\in \mathbb{R}^N$, energetic channel $\bm g^{t=0}\in \mathbb{R}^P$, initial reinforcement $\rho_0=0$, and $\bm w_{\rm best}\leftarrow\sign(\bm{a}^0)$.

        \For{$t=1,\ldots,t_{\max}$}

        \State Perform one rAMP update:
        \begin{align}
            (\bm h^t,\bm a^t,\bm g^t)
            \leftarrow
            \mathrm{rAMPstep}
            \left(
            \bm h^{t-1},\bm a^{t-1},\bm g^{t-1};
            \rho_{t-1},\beta
            \right). \hfill &  & \text{See Algorithm~\ref{alg:ramp_step} for the explicit update.}
        \end{align}

        \State Update the reinforcement:
        \begin{equation}
            \rho_t \leftarrow 1-(1-\rho)^t .
        \end{equation}

        \State Construct the binary estimator:
        \begin{equation}
            \bm w^t \leftarrow \sign(\bm a^t).
        \end{equation}

        \If{$\epsilon(\bm w^t)<\epsilon(\bm w_{\rm best})$}
        \State $\bm w_{\rm best}\leftarrow\bm w^t$.
        \EndIf

        \If{$\epsilon(\bm w^t)\leq\epsilon^{\rm targ}$}
        \State \Return success, $\bm w^t$.
        \EndIf

        \EndFor

        \State \Return failure, $\bm w_{\rm best}$.

    \end{algorithmic}
\end{algorithm}

\begin{figure}[H]
    \centering
    \includegraphics[width=0.5\linewidth]{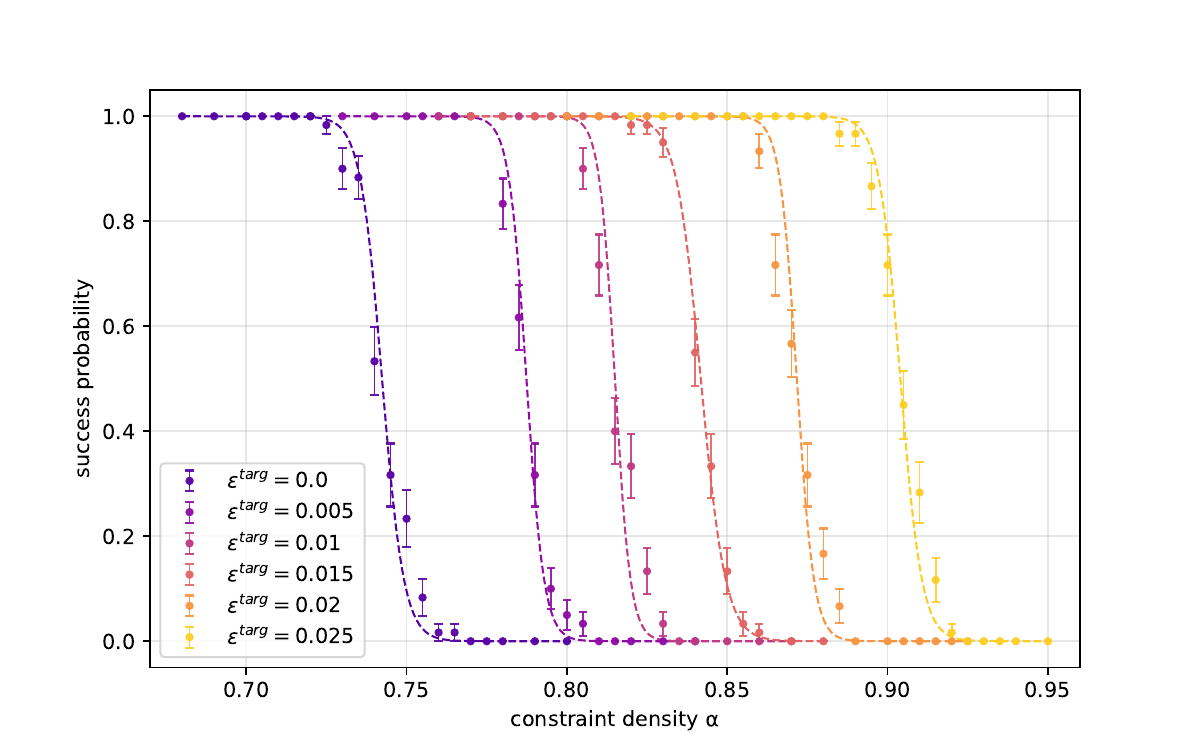}\hfill
    \includegraphics[width=0.5\linewidth]{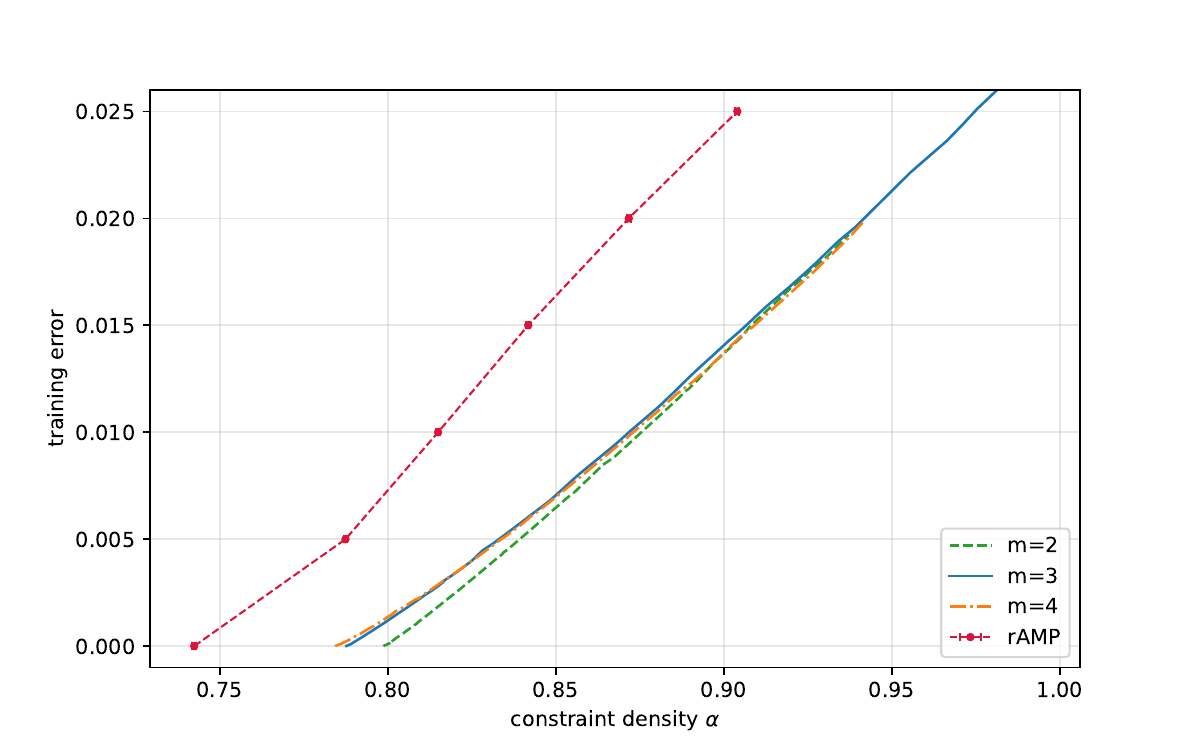}
    \caption{OGP vs rAMP algorithm in the ABP, storage setting. Left panel: probability of finding a configuration with target training error $\epsilon^{\rm targ}$ vs $\alpha$ using the rAMP algorithm at fixed temperature as given in~\ref{alg:ramp_fixed_beta}. Here we have used $N=8000$ and we have averaged the results over $60$ samples. The dashed lines represent sigmoidal fits to the data. Right panel: phase diagram. The red points represent the performance achieved by rAMP at finite temperature and for $N=8000$ and averaged over $60$ samples. For each point we have optimized the reinforcement parameter $\rho$ and the inverse temperature $\beta$ in order to achieve the highest value of the constraint density possible. The green, blue, and yellow curves denote the boundaries marking the presence of $m$-OGP for $m=2$, $3$, and $4$, respectively.
    }
    \label{fig:algorithmic_phase_diagram}
\end{figure}

To turn the AMP marginals into an actual binary assignment with low training error, we use the reinforced AMP (rAMP) algorithm~\cite{Braunstein_2006, Baldassi_2015}. The idea of reinforcement is to add a self-aligning contribution to the AMP field update, which progressively polarizes the local fields and drives the magnetizations $a_i$ towards the vertices of the hypercube. More precisely, if $h_{i,\mathrm{AMP}}^t$ denotes the usual AMP update at iteration $t$ (see equation~\eqref{eq::AMP_local_field} in the Appendix for its expression), we replace it by adding a memory term $\rho_{t-1} h_i^{t-1}$:
\begin{equation}
    h_i^t = h_{i,\mathrm{AMP}}^t + \rho_{t-1}h_i^{t-1}.
\end{equation}
Usually the reinforcement strength $\rho_{t}$ is set to zero at time $t=0$ and increased during the dynamics. We have used a schedule of the type
\begin{equation}
    \rho_t = 1-(1-\rho)^t ,
    \label{eq:reinforcement_schedule_main}
\end{equation}
where $\rho$ controls the rate at which reinforcement is switched on. If
$\rho=0$ one has $\rho_t=0$ for all $t$, and the algorithm reduces to
standard AMP.

The complete procedure is summarized in Algorithm~\ref{alg:ramp_fixed_beta}, and detailed in Appendix~\ref{app:ramp_algorithm}. At each iteration we perform one rAMP update at the current value of $\beta$, construct the binary estimator $\bm w^t=\sign(\bm a^t)$, and measure its training error. The run is declared successful if $\epsilon(\bm w^t)\leq\epsilon^{\rm targ}$ before the maximal number of iterations is reached. We also record the smallest training error encountered along the trajectory.

The left panel of Figure~\ref{fig:algorithmic_phase_diagram} shows the empirical success probability as a function of $\alpha$ for several values of the target training error, in the storage setting and for $N=8000$. For each $\epsilon^{\rm targ}$, the success probability has a sharp drop as $\alpha$ is increased. As expected, allowing a larger training error shifts this algorithmic transition to larger values of $\alpha$: low-energy states with positive error remain accessible in a range of densities where exact solutions are no longer found by the algorithm. For each target training error the algorithmic transition can be estimated by fitting the data to a sigmoid and finding the point where the success probability is $1/2$. In the right panel of Fig.~\ref{fig:algorithmic_phase_diagram} we summarize the same data in a phase diagram. The red points show the best rAMP algorithmic threshold for each considered target training error $\epsilon^{\rm targ}$. Those points have been obtained by tuning the rAMP hyperparameters, namely the reinforcement $\rho$ and the inverse temperature $\beta$, see Appendix~\ref{app:ramp_algorithm} for additional details. When $\epsilon^{\rm targ}=0$, we set $\beta=\infty$ and the procedure reduces to the standard zero temperature rAMP algorithm~\cite{Braunstein_2006}. Those points are to be compared to the OGP thresholds which we present for $m=2$, 3 and 4. 
The rAMP algorithm qualitatively follows the trend of the OGP thresholds.

\begin{figure}[b]
    \centering
    \includegraphics[width=0.45\linewidth]{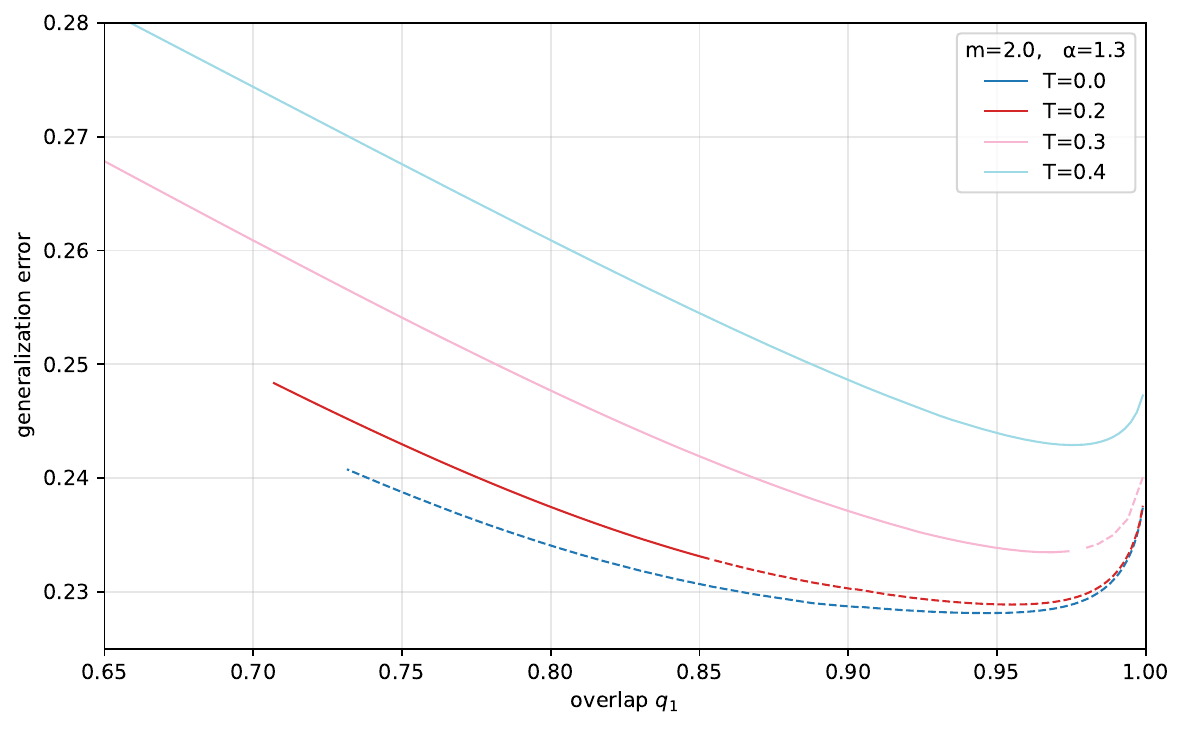}
    \hfill
    \includegraphics[width=0.45\linewidth]{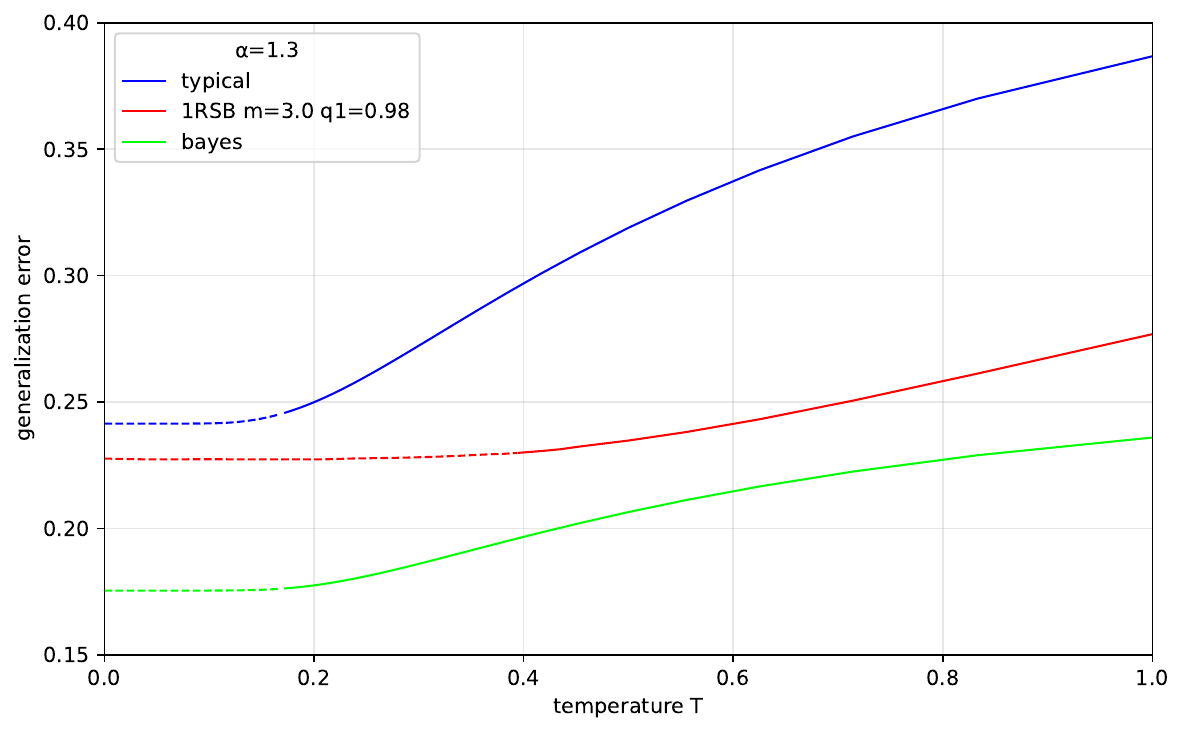}
    \caption{Left panel: generalization of a clone drawn from the constrained measure with $m=2$~\eqref{eq::cloned_partition} versus the mutual clone overlap $q_1$, for several values of the inverse temperature $\beta$. Here we set $\alpha=1.3$, corresponding to a constraint density in the hard region; the qualitative behavior, however, is unchanged for other values of $\alpha$. The plot shows that for each inverse temperature there exists an overlap $q_1^\star$ with minimal generalization error. Right panel: the red line represents the generalization of a clone drawn from the constrained measure with $m=3$ and $q_1=0.98$ versus the temperature. We also display the typical generalization error (blue curve, obtained from the single-clone Gibbs measure~\eqref{eq::uncloned_partition_function}) and the {Bayesian} generalization error (green) for comparison. They are defined in Appendix \ref{app:sec_gen_err}. In both panels, the dashed segments of the curves correspond to non-physical branches with negative entropy. }
    \label{fig:TS_generalization_anal}
\end{figure}

\section{Following wide minima in temperature in the teacher--student framework}
\label{sec:following}

\begin{figure}[b]
    \centering

    \includegraphics[width=0.32\linewidth]{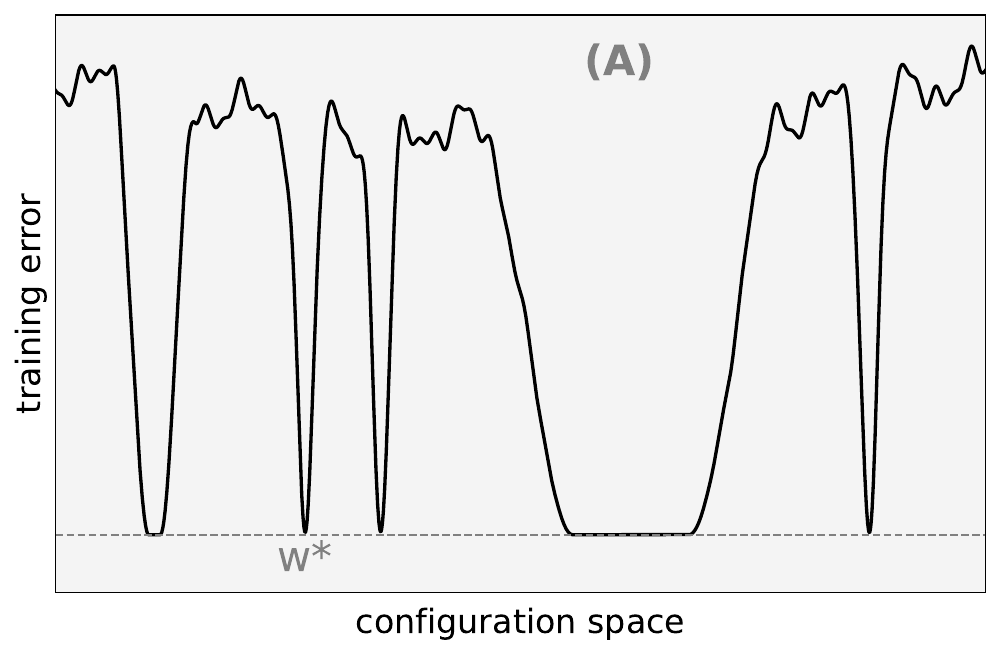}\hfill
    \includegraphics[width=0.31\linewidth]{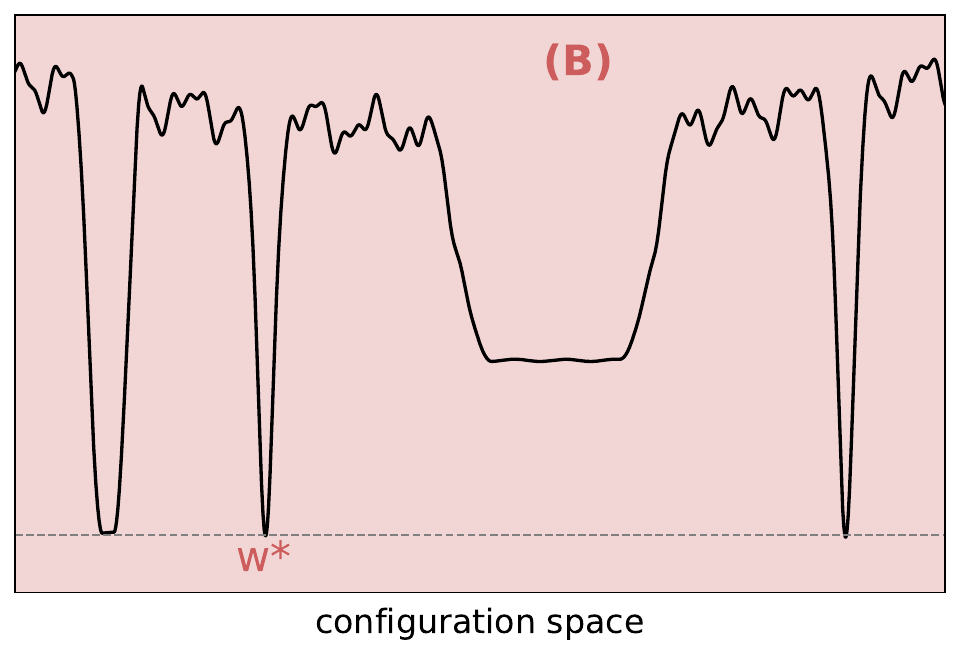}\hfill
    \includegraphics[width=0.31\linewidth]{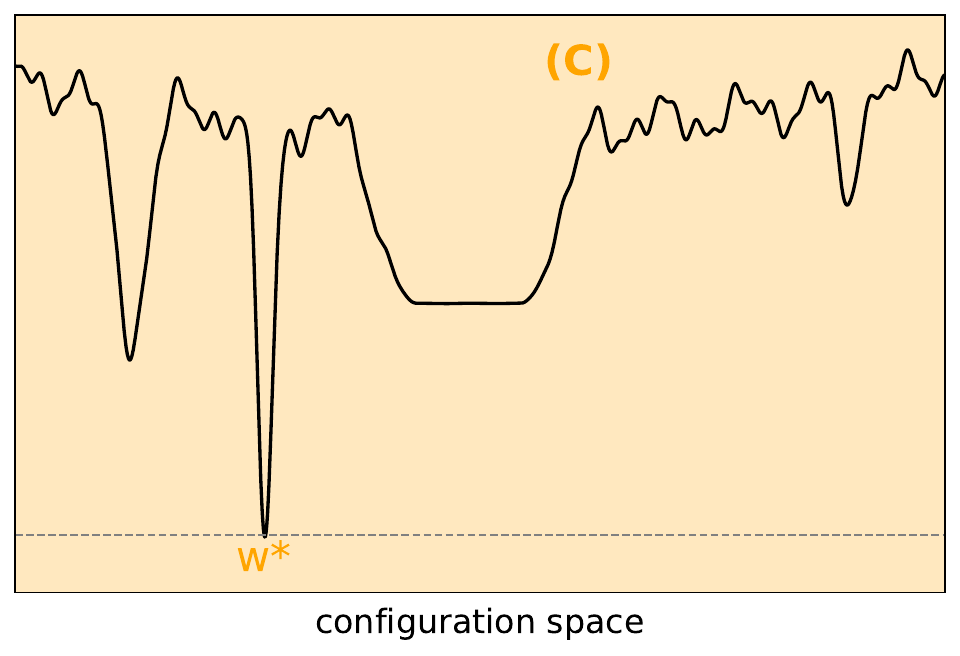}
    \vspace{0.8em}
    \includegraphics[width=0.7\linewidth]{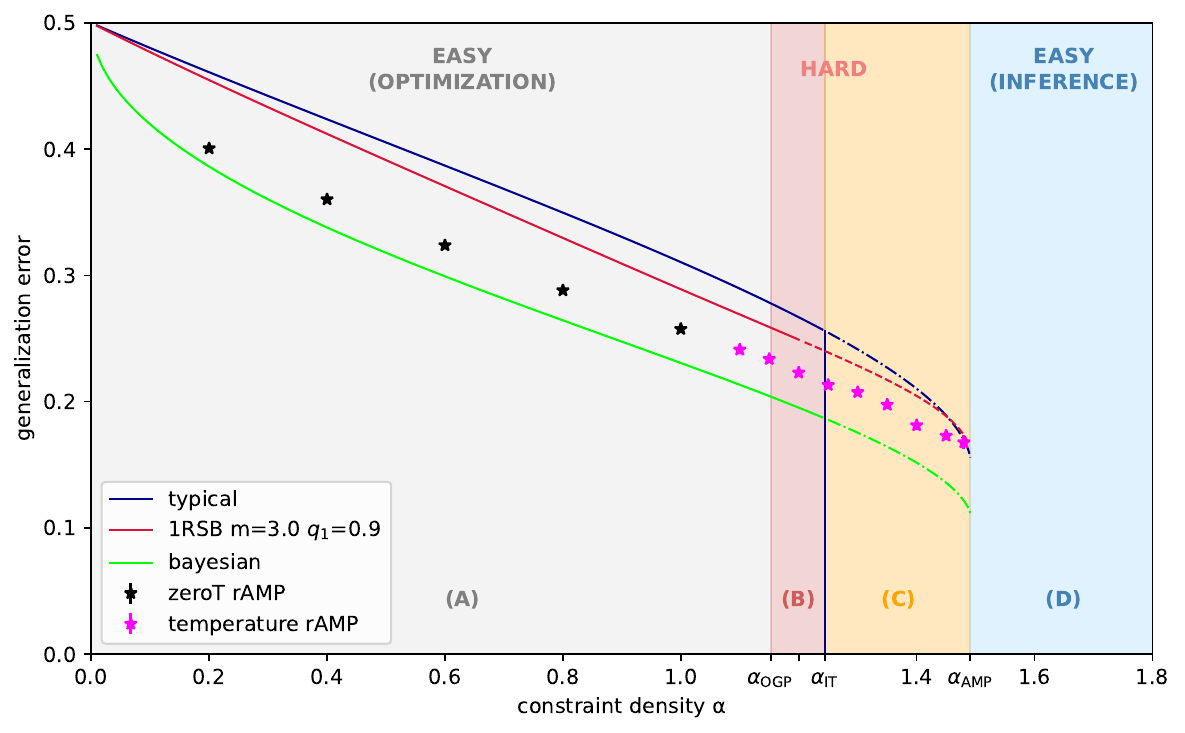}
    \caption{Schematic phase diagram of the binary teacher--student perceptron. The upper panels illustrate the structure of the training error landscape in the different regimes, while the lower panel shows the corresponding generalization error as a function of the constraint density $\alpha$. For $\alpha<\alpha_{\rm OGP}\simeq 1.154$, a wide minimum of zero-training error solutions distinct from the teacher $\bm w^\star$ coexists with smaller, isolated clusters of solutions. The wide minimum is easily accessible to local algorithms (such as the zero temperature rAMP~\cite{Braunstein_2006}, black stars in the bottom panel), whereas the isolated clusters are not. This makes the task of optimization, i.e. of finding a zero training error configuration easy (phase A in the diagram). Nevertheless, exact inference remains information-theoretically impossible below $\alpha_{\rm IT}\simeq 1.249$. For $\alpha>\alpha_{\rm OGP}$, the wide minimum moves to positive training error and therefore no longer contains solutions. In the regime $\alpha_{\rm OGP}<\alpha<\alpha_{\rm IT}$ (phase B), zero-training error configurations distinct from the teacher survive only in isolated clusters separated by an overlap gap and are thus inaccessible to stable algorithms. For $\alpha_{\rm IT}<\alpha<\alpha_{\rm AMP}\simeq 1.492$ (phase C), the only zero-training error configuration is the teacher, which nevertheless remains algorithmically inaccessible. However, the wide minimum of configurations at positive training error can still be targeted by the finite temperature rAMP algorithm described in~\ref{alg:ramp_fixed_beta} (fuchsia stars) and retain good generalization capabilities. Finally, for $\alpha>\alpha_{\rm AMP}$ (phase D), the teacher can be efficiently recovered. We also show in the bottom panel the generalization error of typical (isolated) solutions (blue), of a clone drawn from the constrained measure~\eqref{eq::cloned_partition} with $m=3$ and $q_1=0.9$ (red). The green curve represents the Bayesian error i.e. the error of the barycenter over all typical solutions. The dashed parts of the curves correspond to an non-physical negative entropy.}
    \label{fig:generalization}
\end{figure}

We now turn to the teacher--student setting and investigate the generalization properties of finite-energy configurations that can be targeted by algorithms.

As shown in the previous section, the wide and flat region of zero training error solutions fractures once $\alpha$ exceeds $\alpha_{\rm OGP}\simeq 1.154$ due to the presence of OGP in the space of solutions. Nevertheless, robust dense configurations continue to exist at positive training error and may, in principle, be targeted by algorithms operating at finite temperature. In this section, we study the question of whether those regions at finite training error retain good generalization properties. We first show analytical evidence that such configurations have rather good generalization and then provide numerical evidence that these regions are also algorithmically accessible.

Our analytical approach is based on a comparison between two classes of estimators. The first consists of typical students sampled from the Gibbs measure in Eq.~\eqref{eq::uncloned_partition_function}. As discussed above, this measure is dominated by isolated frozen configurations, which may have a nonzero overlap with the teacher but are not expected to be efficiently accessible to stable algorithms. The second class is obtained from the constrained $m$-clone measure introduced in Eq.~\eqref{eq::cloned_partition}, in which $m$ students are constrained to have mutual overlap $q_1$. By forcing several configurations to remain close to one another, this measure favors regions with large local entropy and therefore provides a proxy for wide and flat portions of the landscape.

The left panel of Figure~\ref{fig:TS_generalization_anal} shows the generalization error of a clone drawn from the constrained measure as a function of the imposed mutual overlap $q_1$. At fixed $\alpha$ and $\beta$, the dependence on $q_1$ is non-monotonic. For small $q_1$, the clones are too far apart to identify a coherent local structure, whereas the limit $q_1\to 1$ approaches an isolated configuration. Between these two limits, the generalization error reaches a minimum at an optimal overlap $q_1^\star$. 
The same analysis can be repeated at finite temperature. Increasing the temperature generally worsens generalization, as expected, but the degradation is gradual. Wide regions can therefore be continuously followed from zero to positive temperature while retaining a generalization error substantially smaller than that of typical configurations. Searching for robust configurations with low, but nonzero, training error therefore does not immediately compromise their alignment with the teacher. This behavior is shown more directly in the right panel of Fig.~\ref{fig:TS_generalization_anal}, where we compare the optimal generalization error of the constrained clones with that of a typical Gibbs configuration and with the Bayesian estimator associated with the barycenter of the Gibbs measure (see Appendix \ref{app:sec_gen_err}). Over the temperature range shown, the constrained-clone estimator continues to outperform a typical isolated configuration. 

We now place this result in the phase diagram of the binary teacher--student perceptron and examine whether these informative wide regions are also algorithmically accessible. Figure~\ref{fig:generalization} summarizes the resulting picture. For $\alpha<\alpha_{\rm OGP}\simeq 1.154$, a wide minimum containing zero-training error configurations coexists with smaller, isolated clusters. The wide minimum is accessible to local algorithms, whereas the isolated clusters are not. Once $\alpha$ exceeds $\alpha_{\rm OGP}$, the wide minimum moves to positive training error. In the interval
$\alpha_{\rm OGP}<\alpha<\alpha_{\rm IT}\simeq 1.249$, exact solutions distinct from the teacher survive only in isolated clusters separated by an overlap gap. For
$\alpha_{\rm IT}<\alpha<\alpha_{\rm AMP}\simeq 1.492$, the teacher is the only zero-training error configuration in the large $N$ limit, but it remains inaccessible to stable algorithms. Nevertheless, the finite-training error continuation of the wide minimum persists throughout this hard region.

The finite-temperature rAMP algorithm introduced in the previous section is able to target this positive-energy wide region, as shown by the fuchsia symbols in Fig.~\ref{fig:generalization}. The figure therefore combines two complementary conclusions: the analytical calculation of Figure ~\ref{fig:TS_generalization_anal} shows that wide finite-energy configurations retain favorable generalization properties, while the numerical results of Fig.~\ref{fig:generalization} show that such configurations can be reached algorithmically beyond the zero-temperature OGP threshold. Finally, for $\alpha>\alpha_{\rm AMP}$, the teacher itself becomes efficiently recoverable.

Finite temperature thus plays a dual role. It relaxes the exact-fitting constraint and restores access to a wide basin in the hard phase, while preserving much of the statistical information carried by the corresponding zero-temperature dense region.

\section{Conclusions}
We studied the finite-temperature geometry of the solution space of binary perceptron models, extending the zero-temperature picture of frozen 1RSB structure and the overlap gap property to the regime where imperfect classification is allowed and statistically weighted. We found that the equilibrium measure remains dynamically frozen at every finite temperature, and traced this to a discontinuity of the single-pattern Gibbs weight at the decision boundary. Smoothing the Gibbs weight, as in Horner's construction~\cite{Horner-storage}, removes the frozen solution at any positive temperature, while using the log-potential on constraints satisfied with high margin, as in ~\cite{straziota2026generative}, removes freezing down to zero temperature. At the level of atypical dense regions, we showed that the OGP threshold can be continuously followed in temperature, growing as $\alpha_{\rm OGP}(T)$, consistent with the intuition that tolerating errors makes room for more constraints - a trend qualitatively reproduced by a finite-temperature message-passing algorithm. In the teacher-student setting, these finite-energy dense regions remain informative about the planted signal, and thermal noise extends the range of constraint densities over which good generalization is algorithmically achievable, even when exact recovery of the teacher is information theoretically impossible.

An open question for future work is whether these finite-temperature wide minima are responsible for the information-theoretic hardness of exact inference in the regime $\alpha_{\rm IT}<\alpha<\alpha_{\rm AMP}$. It also remains to be understood whether, for $\alpha>\alpha_{\rm AMP}$, the teacher becomes embedded within such wide finite-energy minima, thereby explaining the empirical success of replicated or annealed search strategies.

\textbf{Acknowledgments}.
We thank Gianmarco Perrupato for interesting discussions.

\bibliographystyle{apsrev4-1}
\bibliography{biblio}

\appendix

\startcontents[appendices]

\section*{Appendices}
\printcontents[appendices]{}{1}{%
    \setcounter{tocdepth}{2}%
}

\section{The free entropy} \label{sec::free_entropy}

The quenched average in the definition of the free entropy in equation~\eqref{eq::free_entropy} of the main text can be performed introducing $n$ \emph{virtual} replicas of the system and using the replica trick:
\begin{equation}
    \phi = \lim_{N\to\infty} \lim_{n\to 0} \frac{\langle Z^n_\mathcal{D} \rangle_{\mathcal{D}} - 1 }{nN}  \,.
\end{equation}
After standard manipulations, one finds the general expression for the replicated partition function:
\begin{equation}
    \langle Z^n_\mathcal{D} \rangle_{\mathcal{D}} =\int \prod_{a<b} dq_{ab} d\hat{q}_{ab} \prod_{a} dr_a d\hat{r} _a \, e^{NS(\bm{q}, \bm{\hat{q}}, \bm{r}, \bm{\hat{r}})} \,,
\end{equation}
where $q_{ab}$ and $r_a$ physically represent the student-student overlap matrix and the teacher-student overlap -- when a teacher is present, $r_a > 0$, see Table \ref{tab:models} -- respectively \cite{malatesta2023high}. The function $S$ is decomposed into an \emph{entropic} $G_S$ and an \emph{energetic} term $G_E$ as:
\begin{subequations}
    \label{eq::starting_formulas}
    \begin{align}
        S(\bm{q}, \bm{\hat{q}}, \bm{r}, \bm{\hat{r}})   & = G_{S}(\bm{q}, \bm{\hat{q}}, \bm{r}, \bm{\hat{r}}) + \alpha G_E(\bm{q}, \bm{r}) \,,                                                                                               \\
        G_S(\bm{q}, \bm{\hat{q}}, \bm{r}, \bm{\hat{r}}) & = -\frac{1}{2} \sum_{a\ne b}q_{ab} \hat{q}_{ab} -\sum_{a}r_a\hat{r}_a + \ln \sum_{\{w^{a}=\pm 1\}}\ e^{\frac{1}{2}\sum_{a\ne b}\hat{q}_{ab}w^{a}w^{b}+ \sum_a  \hat{r}_a w^a } \,, \\
        G_{E}(\bm{q}, \bm{r})                           & = \ln \int \frac{d\nu d\hat{\nu}}{2\pi} \prod_{a} \frac{du_{a}d\hat{u}_{a}}{2\pi} \prod_{a} \mathcal{K} \left(\sign(\nu) u_{a}\right)
        e^{i\sum_{a} u_{a}\hat{u}_{a} + i \nu\hat{\nu} -\frac{1}{2}\sum_{ab}q_{ab}\hat{u}_{a}\hat{u}_{b} - \frac{\hat{\nu}^2}{2} -\hat{\nu}\sum_a \hat{u}_{a} r_a }\,.
    \end{align}
\end{subequations}
The free entropy $\phi$ can be computed by solving the following extremization problem
\begin{equation}
    \label{appx:limS}
    \phi = \lim_{n \to 0} \mathrm{extr}_{\boldsymbol{q}, \hat{\boldsymbol{q}}, \boldsymbol{r}, \hat{\boldsymbol{r}}} \, \frac{S(\boldsymbol{q}, \hat{\boldsymbol{q}}, \boldsymbol{r}, \hat{\boldsymbol{r}})}{n} \,.
\end{equation}
Note how the energetic term depends on the generic Gibbs weight $\mathcal{K}$; see Eqs. \eqref{eq:K_ABP_main}, \eqref{eq:K_SBP_main} in the main text. We keep the discussion general, but for clarity of the reader, Table \ref{tab:models} contains a summary of how to specify the equations for the model considered in this paper (ABP/SBP) and setting (storage/teacher-student).

\begin{table}[H]
    \centering
    \begin{tabular}{|c|c|c|}
        \hline
        \; SBP (with margin $\kappa > 0$) \; &
        \; \; ABP (storage) \;\;             &
        \; \; ABP (teacher-student) \;\;                                                                      \\
        \hline
        \; \; $\mathcal{K}^{SBP}$ \; \;      & $\mathcal{K}^{ABP}$ & $\mathcal{K}^{ABP}$                      \\
        \; \; $r = \hat{r}= 0$ \; \;         & $r = \hat{r}= 0$    & $r = r^\star, \; \hat{r}= \hat{r}^\star$ \\
        \hline
    \end{tabular}
    \caption{Recipe for each model, to be inserted in Eq. \eqref{eq::starting_formulas}. The single-pattern Gibbs weight for the ABP and SBP models, $\mathcal{K}_{ABP}$ and  $\mathcal{K}_{SBP}$, are defined in Eqs. \eqref{eq:K_ABP_main} and \eqref{eq:K_SBP_main}, respectively.
    }
    \label{tab:models}
\end{table}

\subsection{RS Ansatz} \label{sec::RS_entropy}

We consider here the standard replica symmetric (RS) assumption on the student overlap matrix and its conjugate
\begin{align*}
    q_{ab}      & = \delta_{ab} + (1-\delta_{ab}) q\, \\
    \hat q_{ab} & = (1-\delta_{ab}) \hat q\,;
\end{align*}
moreover we impose for the teacher-student overlaps
\begin{align*}
    r_a      & = r\,       \\
    \hat r_a & = \hat r\,.
\end{align*}
A standard computation gives the free entropy \eqref{appx:limS}:
\begin{subequations}
    \label{eq::rs-starting_formulas}
    \begin{align}
        S^{\mathrm{RS}}(q, \hat{q}, r, \hat{r})             & = \mathcal{G}^{\mathrm{RS}}_S(q, \hat{q}, r, \hat{r}) + \alpha \mathcal{G}^{\mathrm{RS}}_{E}(q, r)\,,     \\
        \mathcal{G}^{\mathrm{RS}}_S(q, \hat{q}, r, \hat{r}) & =  -\frac{\hat{q}}{2}(1-q) -\hat{r}r + \int Dz\, \ln 2\cosh\left(\sqrt{\hat{q}}z + \hat{r}\right)\,,      \\
        \mathcal{G}^{\mathrm{RS}}_{E}(q, r)                 & = 2\int Dz\, H\left(-\frac{rz}{\sqrt{q-r^2}} \right)\ln \mathcal{H} \left(\sqrt{q}z, \sqrt{1-q}\right)\,,
    \end{align}
\end{subequations}
where
\begin{equation}
    \label{eq::HbetaKernel}
    \mathcal{H}(x, y) = \int Dz \, \mathcal{K}(x + y z)\,.
\end{equation}
In particular for the ABP and SBP with standard training error loss function one has
\begin{subequations}
    \begin{align}
        \mathcal{H}(x, y) & = e^{-\beta} + (1-e^{-\beta}) \mathcal{H}_{\infty}(x, y) \,, \label{eq:Hbeta}
    \end{align}
\end{subequations}
where $\mathcal{H}_{\infty}(x, y)$ depends on the particular model we consider:
\begin{subequations}
    \begin{align}
        \mathcal{H}_{\infty}^{\mathrm{ABP}}\left(x, y\right)                  & = H\left(-\frac{x}{y}\right)\,,                            \\
        \label{eq:Hsbp} \mathcal{H}_{\infty}^{\mathrm{SBP}}\left( x, y\right) & = \sum_{s=\pm 1} s H\left(\frac{-s\kappa + x}{y}\right)\,,
    \end{align}
\end{subequations}
being $H(x) = \int_x^{\infty} Dh = \frac{1}{2} \mathrm{Erfc}\left(\frac{x}{\sqrt{2}}\right)$. The thermodynamic free entropy is computed at the values of $q, \hat{q}, r, \hat{r}$ extremizing the function $S^{RS}$:
\begin{equation}
    \phi^{\mathrm{RS}} = \mathrm{extr}_{q, \hat{q}, r, \hat{r}} \left[ S^{\mathrm{RS}}(q, \hat{q}, r, \hat{r}) \right]\,.
\end{equation}
In the SBP the equilibrium value of the overlap is $q = \hat{q} = 0$ by symmetry\footnote{The same argument for the SBP applies for the parameter $q_0$ in the 1RSB computation.}.

\subsection{1RSB Ansatz and entropy of the \emph{m}-cloned system} \label{sec::1RSB_entropy}

\begin{figure}[t]
    \centering
    \includegraphics[width=0.5\linewidth]{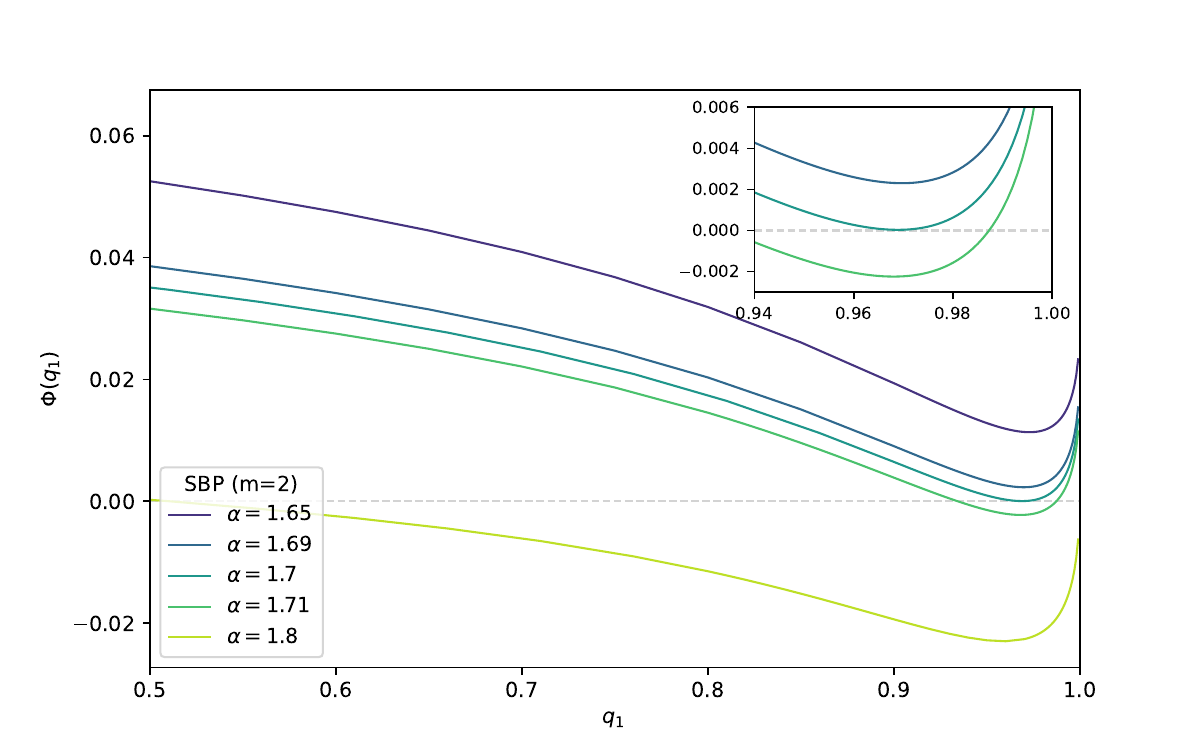}\hfill
    \includegraphics[width=0.5\linewidth]{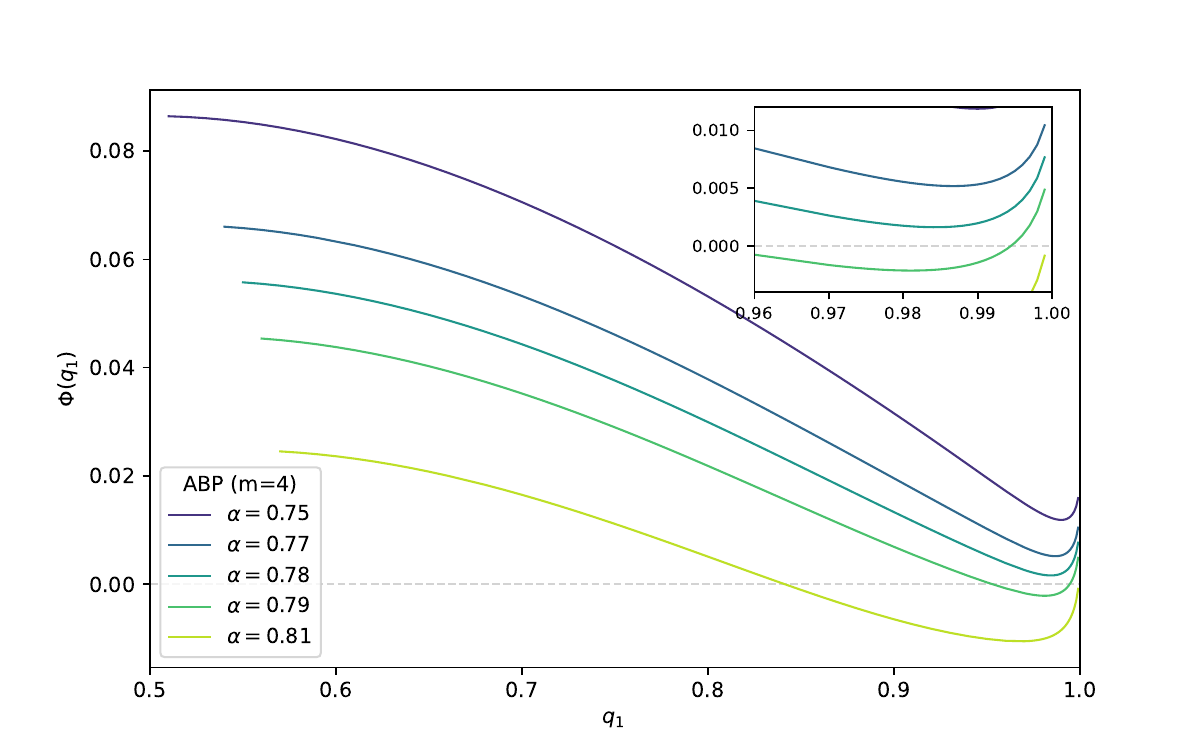}\hfill
    \caption{Left panel: SBP: $\phi_m(q_1)$ for $\kappa=1$, $m=2$, $\beta \to \infty$ for different values of $\alpha$. At $\alpha \simeq 1.700$, the minimum of the free entropy reaches zero, marking the OGP threshold, see Table \ref{tab:sbp_ogp_comparison}.
        Right panel: ABP: $\phi_m(q_1)$ for $m=4$ and $\beta \to \infty$. Our OGP threshold estimate for the ABP is $\alpha_{\rm OGP}\simeq 0.784$, see Figure \ref{fig:alphaOGP_SBP_and_ABP_Storage}.}
    \label{fig:phi(q1)}
\end{figure}

\begin{figure}[t]
    \centering
    \includegraphics[width=0.5\linewidth]{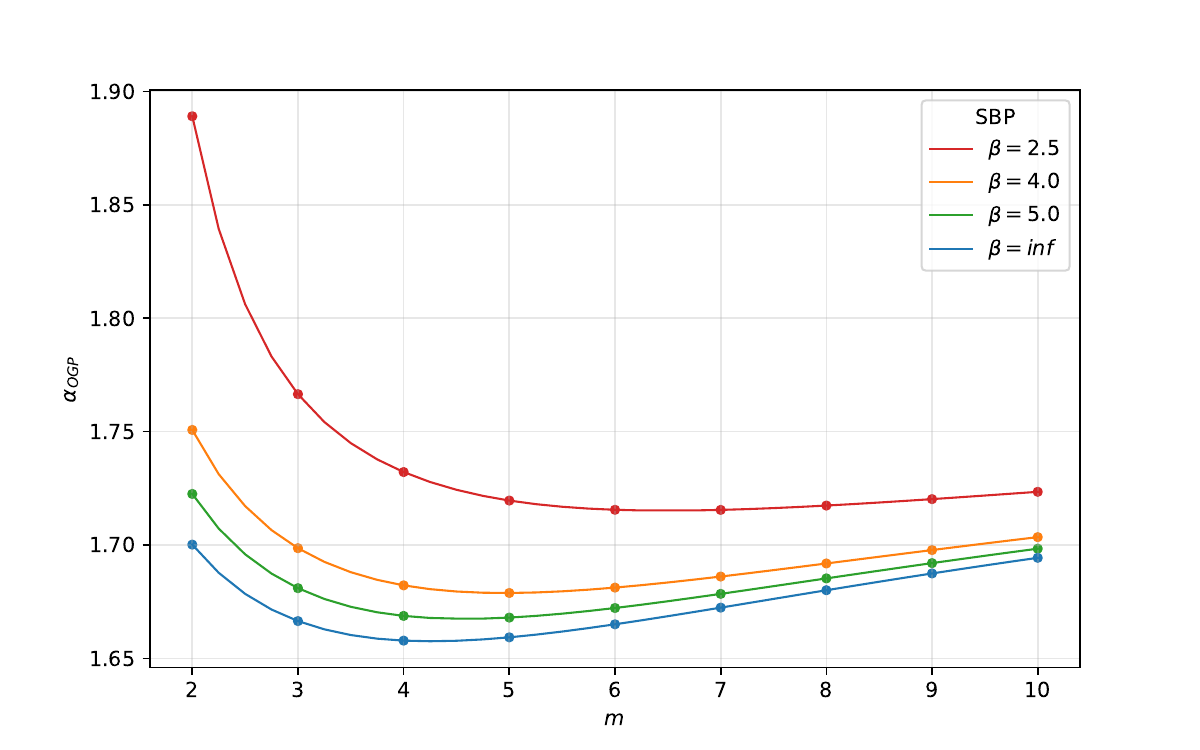}\hfill
    \includegraphics[width=0.5\linewidth]{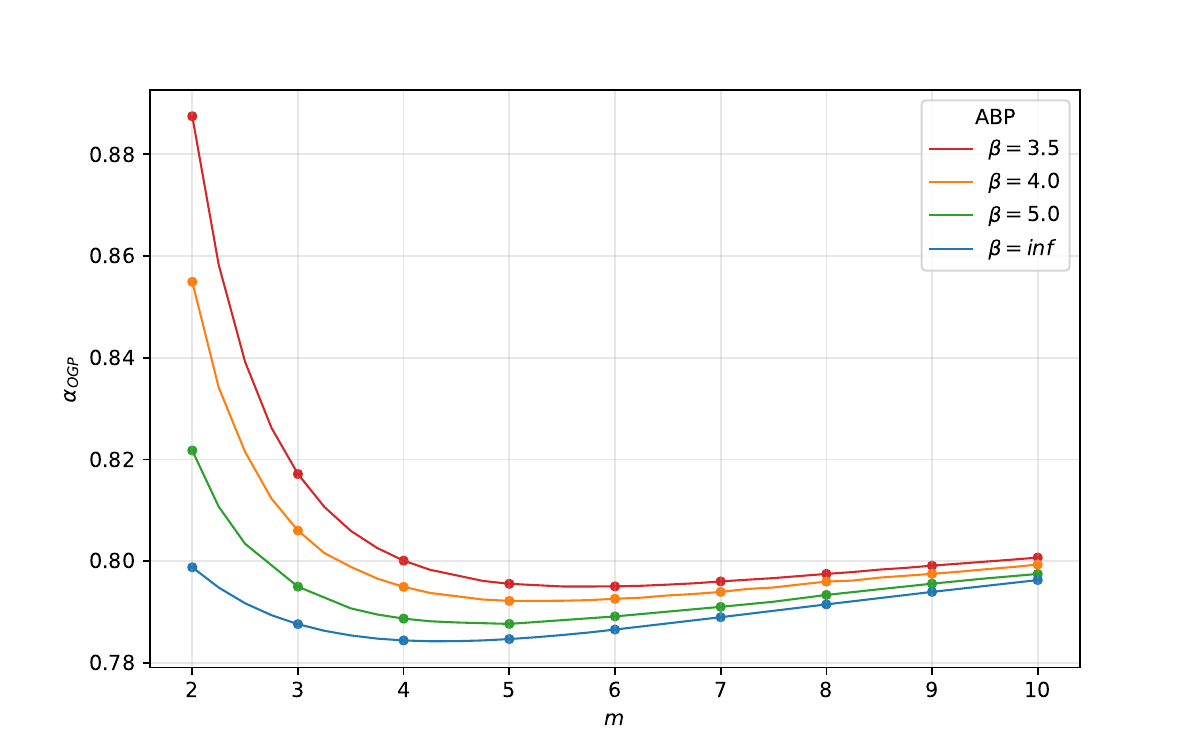}\hfill

    \caption{$\alpha_{\rm OGP}$ as a function of $m$ for the SBP with $\kappa = 1$ and ABP (right panel). At larger temperatures, the curves become less non-physical as the increasing part of the curves become less pronounced. Therefore the RS approximation we made on the cloned  free entropy in equation~\eqref{eq::free_entropy_clones} becomes less dramatic.}
    \label{fig:alphaOGP_SBP_and_ABP_Storage}
\end{figure}

A more refined parameterization of the student overlap matrix $q_{ab}$ is in general needed to compute the equilibrium value of the free entropy~\cite{parisi1979toward}. We consider here a 1-step replica symmetry breaking (1RSB) Ansatz which consists in imposing
\begin{subequations} \label{eq:q_ab_1rsb}
    \begin{align}
        q_{ab}      & = q_0 + (q_1-q_0) I_{ab}^{(n,m)} + (1-q_1) I_{ab}^{(n,1)}                                                    \\
        \hat q_{ab} & = \begin{cases}\hat q_0 + (\hat q_1 - \hat q_0) I_{ab}^{(n,m)} + (1-\hat q_1) I_{ab}^{(n,1)} \quad & a\neq b \\
                                     0                                                                                   & a=b
                        \end{cases}
    \end{align}
\end{subequations}
where $I_{ab}^{(n,m)}$ is the $(a,b)$ element of a block matrix of size $n \times n$ whose diagonal blocks have size $m \times m$ and contain all ones and outside of them the matrix is composed of zeros. We leave unchanged the Ansatz over the teacher-student overlaps.\\
The corresponding free entropy is given by:
\begin{subequations}
    \label{eq::1RSB_free_entropy}
    \begin{align}
        \phi^{\mathrm{1RSB}}            & = \mathrm{extr}_{q_0, \hat q_0, q_1, \hat q_1, r, \hat r, m} \left[ S^{\mathrm{1RSB}}(q_0, \hat q_0, q_1, \hat q_1, r, \hat r, m)\right] \label{eq:phi_def}                                                                                                                      \\
        S^{\mathrm{1RSB}}               & = \mathcal{G}^{\mathrm{1RSB}}_S + \alpha \mathcal{G}^{\mathrm{1RSB}}_E \label{eq:S_1rsb}                                                                                                                                                                                         \\
        \mathcal{G}_{S}^{\mathrm{1RSB}} & =-\frac{\hat{q}_{1}}{2}\left(1-q_{1}\right) + \frac{m}{2}\left(q_{0}\hat{q}_{0}-q_{1}\hat{q}_{1}\right) -\hat{r}r + \frac{1}{m}\int Dz_0\,\ln\int Dz_1\left(2\cosh\left(\sqrt{\hat{q}_{0}}z_0+\sqrt{\hat{q}_{1}-\hat{q}_{0}}z_1 + \hat r\right)\right)^{m}\label{eq:Gs_bin-1rsb} \\
        \mathcal{G}_{E}^{\mathrm{1RSB}} & =\frac{2}{m}\int Dz_{0}\,H\left(-\frac{rz_0}{\sqrt{q_0-r^2}}\right)\,\ln\int Dz_{1} \, \mathcal{H}^{m}\left({ \sqrt{q_0}z_{0}+\sqrt{q_1- q_0}z_{1}},{\sqrt{1-q_1}} \right)
        \label{eq:Ge-1rsb}
    \end{align}
\end{subequations}
where $\mathcal{H}$ is defined in \eqref{eq:Hbeta}.\\
Note also that this expression can be used to find the entropy density of $m$-clones of the model constrained to be at a given overlap $q_1$ as in equation~\eqref{eq::free_entropy_clones} \cite{Monasson1995}. Indeed, notice that in this case the $m$ replicas are ``real'', and $n$ virtual replicas of the cloned system are introduced, so that $n$ should be replaced with $nm$ in the previous formulas~\eqref{eq:q_ab_1rsb}. Moreover one should not optimize over both $m$ and $q_1$, as they are treated as external parameters. Mathematically, we have:
\begin{equation}
    \phi_m(q_1) =  \mathrm{extr}_{q_0, \hat q_0, \hat q_1, r, \hat r} \left[ \mathcal{G}^{\mathrm{1RSB}}_S + \alpha \mathcal{G}^{\mathrm{1RSB}}_E\right]\,.
\end{equation}
Figure~\ref{fig:phi(q1)} shows exemplary curves for the SBP and ABP as a function of the overlap $q_1$. As described in the main text, the $m$-OGP threshold at zero temperature can be found by solving \eqref{eq:ogp_zeroT_conditions}:
\begin{align*}
    \phi_m(q_1)                & = 0\,, \\
    \partial_{q_1} \phi_m(q_1) & = 0\,.
\end{align*}
Similar conditions hold for $\alpha_{\rm OGP}$ threshold at $\beta > 0$, see Eqs. \eqref{eq:ogp_finiteE_conditions}.\\
Figure \ref{fig:alphaOGP_SBP_and_ABP_Storage} shows $\alpha_{\rm OGP}$ as a function of $m$ for the SBP and ABP in the storage setting and for different values of the inverse temperature $\beta$. Note that if $m'>m$ one should have $\alpha_{\rm OGP}(m') \le \alpha_{\rm OGP}(m)$, by the very definition of OGP. The increasing parts of those curves are therefore clearly non-physical~\cite{benedetti25}.

Another indication of the non-physicality of such points is given by the computation of their complexity $\Sigma$. Indeed, the complexity of a system, or configurational entropy, represents the log-number of metastable states, that are pure states in which the partition function \eqref{eq::cloned_partition} can be decomposed. Here, it is computed as
\begin{align*}
    \Sigma_m = -m^2\frac{\partial \phi_m}{\partial m}\,,
\end{align*}
following the real replicas computation of \cite{Monasson1995}, with the difference that $q_1$ is set by Eqs. \eqref{eq:ogp_zeroT_conditions}.
Interestingly, it can be verified that the points violating the monotonicity criterion in Fig.  \ref{fig:alphaOGP_SBP_and_ABP_Storage} are also characterized by a negative complexity $\Sigma < 0$, that is non-physical by definition.

We have summarized in table~\ref{tab:sbp_ogp_comparison} the OGP threshold we have obtained for the SBP with $\kappa = 1$ at zero temperature, and we compare them with the rigorous annealed bound established in~\cite{gamarnik2022sbp}.

\begin{table}[H]
    \centering
    \begin{tabular}{|c|c|c|}
        \hline
        \; $m$ \;
            & \; \; RS estimate of $\alpha^{(\mathrm{sbp})}_{\mathrm{OGP}}(m)$ \;\;
            & \; \; Annealed bound~\cite{gamarnik2022sbp} \;\;                                \\
        \hline
        $2$ & $1.7001$                                                              & $1.71$  \\
        $3$ & $1.6664$                                                              & $1.667$ \\
        $4$ & $1.6578$                                                              & --      \\
        $5$ & $1.6593$                                                              & --      \\
        \hline
    \end{tabular}
    \caption{
        OGP threshold for the symmetric binary perceptron with $\kappa=1$: comparison between the RS estimate and the rigorous first moment bounds of Ref.~\cite{gamarnik2022sbp}.  
        Our RS estimates of the OGP thresholds are consistent with the ones reported in~\cite{stojnic2026parametric,stojnic2026ultrametric} obtained using another analytical machinery.}
    \label{tab:sbp_ogp_comparison}
\end{table}

\section{Computation of the dynamical temperature} \label{app:dynamical-temperature}
We carry out here the computation of the dynamical temperature $T_d$. The computation is general for both the storage and teacher-student settings.\\
In order to find out $T_d$, we need to expand the 1RSB free entropy functional $S^{\mathrm{1RSB}}$ given in~\eqref{eq:S_1rsb} around $m=1$ \cite{KirkpatrickThirumalai1987}. At zeroth order we simply get the RS free entropy \eqref{eq::rs-starting_formulas}, with $q=q_0$ and $\hat q = \hat q_0$.
The first order correction is:
\begin{equation}
    S^{\mathrm{1RSB}}(q_0, \hat q_0, q_1, \hat q_1, r, \hat r) = S^{\mathrm{RS}}(q_0, \hat{q}_0, r, \hat r) + (m-1) \tilde \phi(q_0, \hat q_0, q_1, \hat q_1, r, \hat r) + O\left((m-1)^2\right) \,,
\end{equation}
where the function $\tilde \phi$ is defined as
\begin{align}
    \tilde \phi & = \left. \frac{\partial S^{\mathrm{1RSB}}}{\partial m} \right|_{m=1} = \left. \frac{\partial \mathcal{G}_S}{\partial m} \right|_{m=1} + \alpha \left. \frac{\partial \mathcal{G}_E}{\partial m}\right|_{m=1} \,,
\end{align}
and it is given by
\begin{subequations}
    \begin{align}
                      & \tilde \phi(q_0, \hat q_0, q_1, \hat q_1, r, \hat r) = - S^{\mathrm{RS}}(q_0, \hat q_0, r, \hat r) - r \hat r + q_0 \hat q_0  - \frac{\hat q_1}{2} (1+q_1)  + \mathcal{I}_S(\hat q_0, \hat q_1, \hat r) + \alpha \mathcal{I}_E(q_0, q_1, r)                                                                         \\
        \mathcal{I}_S & = e^{-\frac{\hat q_1 - \hat q_0}{2}} \int Dz_0 \, \frac{\int Dz_1 \, \cosh \left(\sqrt{\hat q_0} z_0 + \sqrt{\hat q_1 - \hat q_0} z_1 + \hat r\right) \ln 2\cosh \left(\sqrt{\hat q_0} z_0 + \sqrt{\hat q_1 - \hat q_0} z_1 + \hat r\right) }{\cosh (\sqrt{\hat q_0} z_0+ \hat r)}                                  \\
        \mathcal{I}_E & = 2\int Dz_0 \, H\left(- \frac{r z_0}{\sqrt{q_0 - r^2}}\right) \frac{\int Dz_1 \, \mathcal{H} \left({\sqrt{q_0}z_{0}+\sqrt{q_1-q_0}z_{1}}, {\sqrt{1-q_1}} \right) \ln \mathcal{H} \left({\sqrt{q_0}z_{0}+\sqrt{q_1-q_0}z_{1}}, {\sqrt{1-q_1}} \right)}{\mathcal{H} \left({\sqrt{q_0}z_{0} },{\sqrt{1-q_0}} \right)}
    \end{align}
\end{subequations}
We recall that equilibrium values of the parameters $q_0, \hat{q_0}, q_1, \hat{q_1}, r, \hat r$ are those that extremize $S^{\rm 1RSB}$.
Notice that $q_0$, $\hat{q}_0$, $r$, $\hat r$ in the $m\to 1$ limit need to solve the saddle point equations for the RS free entropy \eqref{eq::rs-starting_formulas}, while $q_1$ and $\hat q_1$ can be found by extremizing $\tilde{\phi}$ only. In summary one has to solve the following saddle point equations:
\begin{subequations}
    \label{eq::SPE_Td}
    \begin{align}
        \partial_{\hat{q}_0} S^{\mathrm{RS}}(q_0, \hat q_0, r, \hat r)             & = 0\,,  \\
        \partial_{q_0} S^{\mathrm{RS}}(q_0, \hat q_0, r, \hat r)                   & = 0\,,  \\
        \partial_{\hat{r}} S^{\mathrm{RS}}(q_0, \hat q_0, r, \hat r)               & = 0\,,  \\
        \partial_{r} S^{\mathrm{RS}}(q_0, \hat q_0, r, \hat r)                     & = 0\,,  \\
        \partial_{\hat{q}_1} \tilde{\phi}(q_0, \hat q_0, q_1, \hat q_1, r, \hat r) & = 0     \\
        \partial_{q_1} \tilde{\phi}(q_0, \hat q_0, q_1, \hat q_1, r, \hat r)       & = 0 \,.
    \end{align}
\end{subequations}
The dynamical temperature $T_d(\alpha)$ for a fixed value of $\alpha$ is defined as the largest temperature for which a solution with $q_1 > q_0$ appears. This can be found numerically as follows. First we numerically solve the first four equations above, respectively for $q_0$, $\hat{q}_0$, $r$ and $\hat r$.
Then consider the last two equations in~\eqref{eq::SPE_Td}. Using the property of the Kernel~\eqref{eq::HbetaKernel} $\partial_y \mathcal{H}(x, y) = y \partial_x^2 \mathcal{H}(x,y)$ and an integration by parts they read
\begin{align}
    \label{eq:fq1a} 0=\frac{\partial \tilde \phi}{\partial \hat q_1} & = - \frac{q_1}{2} + \frac{1}{2}e^{-\frac{\hat q_1 - \hat q_0}{2}} \int Dz_0 \, \frac{\int Dz_1 \, \sinh \left(\sqrt{\hat q_0} z_0 + \sqrt{\hat q_1 - \hat q_0} z_1 + \hat r\right) \tanh \left(\sqrt{\hat q_0} z_0 + \sqrt{\hat q_1 - \hat q_0} z_1 + \hat r\right) }{\cosh (\sqrt{\hat q_0} z_0  + \hat r)}                                                                                        \\
    \label{eq:fqh1a} 0=\frac{\partial \tilde \phi}{\partial q_1}
                                                                     & = - \frac{\hat q_1}{2} + \frac{\alpha}{2} \int   Dz_0 \, \frac{2H\left(-\frac{r z_0}{\sqrt{q_0-r^2}}\right) }  {\mathcal{H}\left( {\sqrt{q_0}z_{0}}, {\sqrt{1-q_0}}\right) }  \int Dz_1 \frac{  \left(\partial_x \mathcal{H}\left( { \sqrt{q_0}z_{0}+\sqrt{q_1-q_0}z_{1}}, {\sqrt{1-q_1}} \right) \right)^2   }{ \mathcal{H}\left( { \sqrt{q_0}z_{0}+\sqrt{q_1-q_0}z_{1}}, {\sqrt{1-q_1}} \right) }
\end{align}
Via equation~\eqref{eq:fqh1a} one expresses $\hat q_1$ in terms of $q_1$ and inserts it in equation~\eqref{eq:fq1a}. One then plots $\partial_{\hat{q}_1} \tilde{\phi}$ as a function of $q_1$. This function will have at any temperature a root when $q_1 = q_0$ corresponding to the RS solution; this is also the only one when $T>T_d(\alpha)$. At $T_d(\alpha)$, $\partial_{\hat{q}_1} \tilde{\phi}$ develops a new root at a value $q_1 > q_0$.

\begin{figure}
    \centering
    \includegraphics[width=0.5\linewidth]{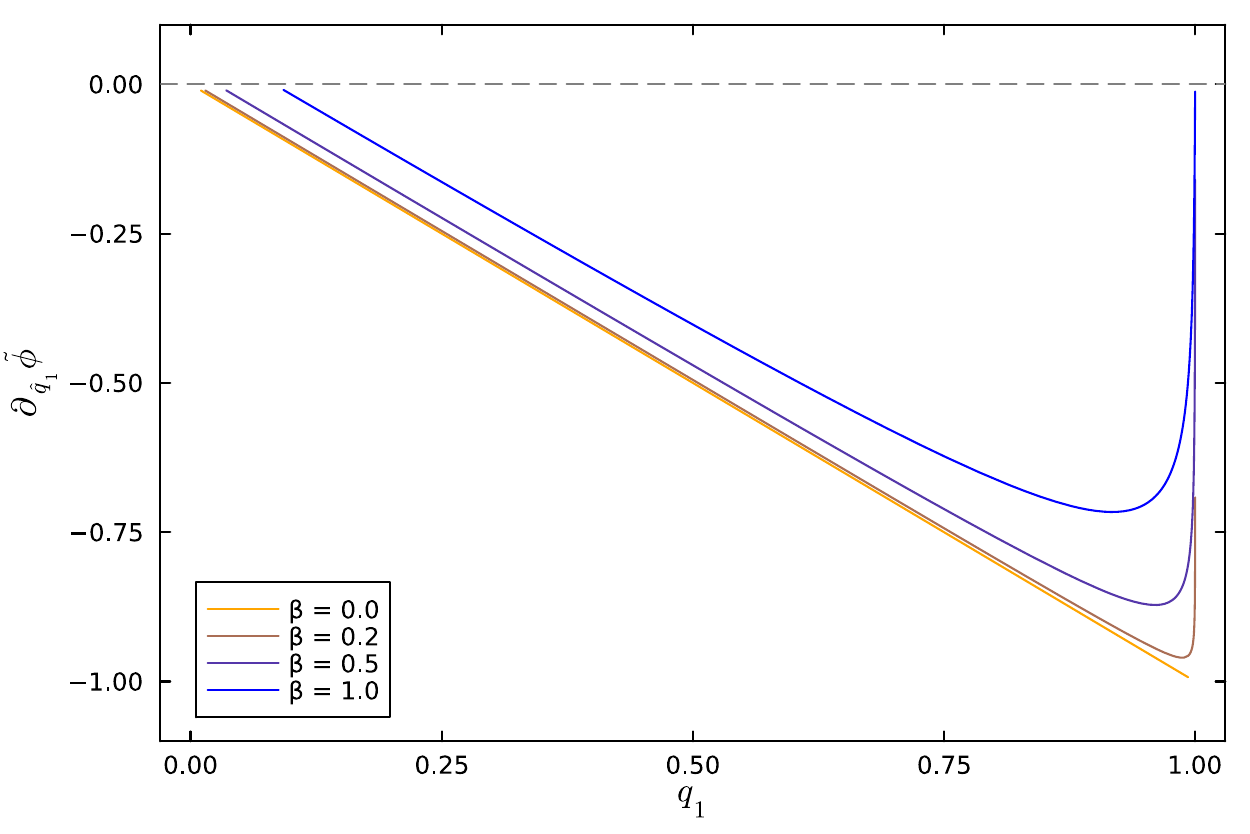}\hfill
    \includegraphics[width=0.5\linewidth]{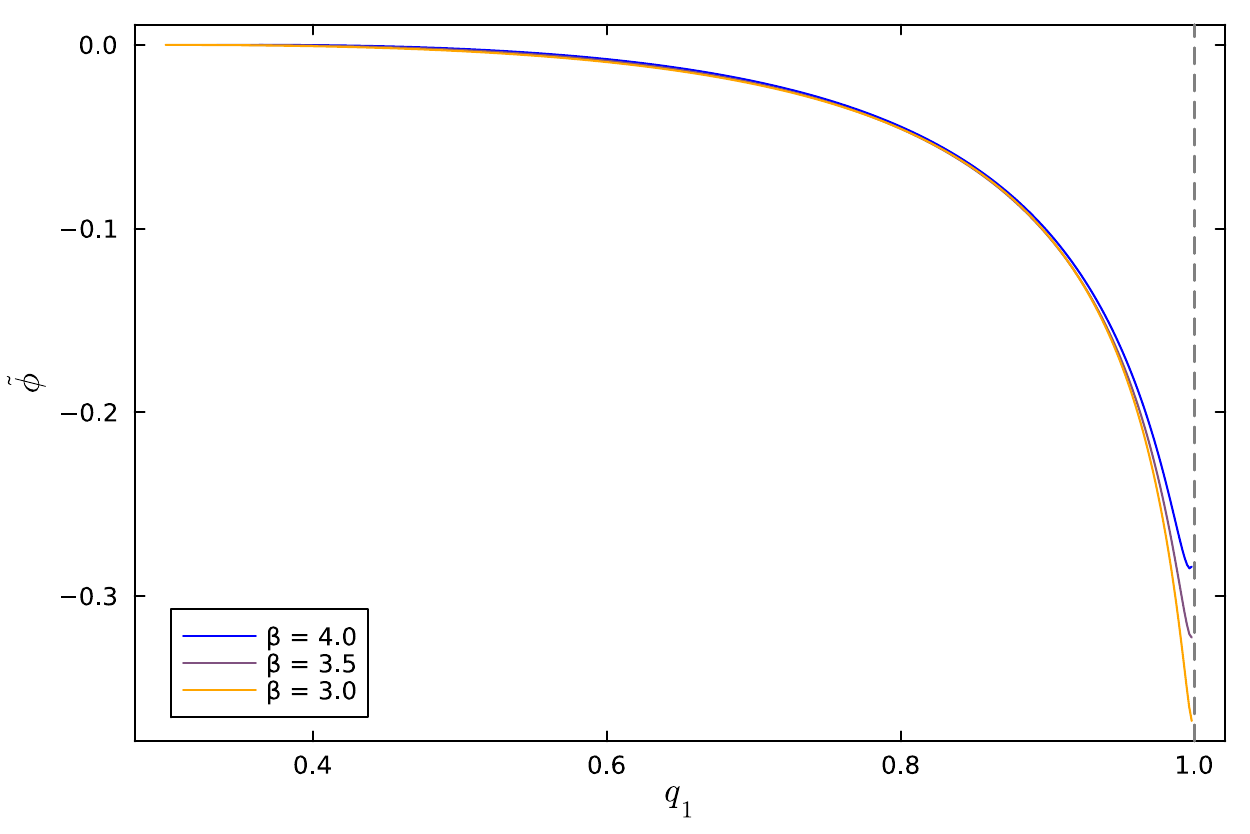}
    \caption{Left: $\partial_{\hat{q}_1} \tilde{\phi}$ for the ABP in the storage case (where $r = \hat r = 0$) as a function of $q_1$, for $\alpha=0.7$ and different values of $\beta$. The solutions of the SPE are the zeros of this function. There is always a zero in $q_1 = q_0$ (the value of $q_0$ depends on $\alpha$ and $\beta$) corresponding to the RS solution. For any $\beta > 0$, another non-trivial solution of the saddle point equation is found in $q_1 = 1$. Right: ABP in the storage setting. The panel presents $\tilde{\phi}$ as a function of $q_1$, for $\alpha=0.7$ and different values of $\beta$: its derivative vanishes at $q_1 = 1$.}
    \label{fig:dynamical_temperature}
\end{figure}

The dynamical temperature depends heavily on the nature of the factor $\mathcal{K}$ which enters into the definition of $\mathcal{H}(x, y)$, see~\eqref{eq::HbetaKernel}. In the following subsection we will specialize to the case of the ABP and SBP constraints and with the standard training error loss, counting the number of unsatisfied constraints. We show that whenever $\beta = 1/T > 0$, the saddle point equations admit the solution $q_1 = 1$.
This can be observed in Figure \ref{fig:dynamical_temperature}, where we plot $\partial_{\hat{q}_1} \tilde{\phi}$ as a function of $q_1$ (for the ABP in the storage case, where one furthermore has $r = \hat r=0$). This means that in the perceptron model the dynamical temperature always diverges in the large $N$ limit.

The solution $q_1=1$ corresponds to the so called \emph{frozen 1RSB} scenario already found in \cite{Kabashima} from the zero-temperature static analysis and is consistent with the results found by Horner~\cite{Horner-storage} in the storage case, which were derived using dynamical mean-field theory.
We will also discuss how to avoid this frozen phase by changing the expression of the energy function.

\subsection{The frozen 1RSB solution}

We specialize here the computation to the ABP for simplicity. Equation \eqref{eq:fqh1a} solved for $\hat{q}_1$ then reads:
\begin{equation}\label{eq:qh1(q1)}
    \hat q_1 
    = \alpha \frac{(1-e^{-\beta})^2}{1-q_1} \int Dz_0 \, \frac{2H\left(-\frac{r z_0}{\sqrt{q_0-r^2}}\right) }  {\mathcal{H}\left( {\sqrt{q_0}z_{0}}, {\sqrt{1-q_0}}\right)}  \int Dz_1  \, \frac{ G\left(\frac{\sqrt{q_0}z_{0}+\sqrt{q_1-q_0}z_{1}}{\sqrt{1-q_1}} \right)^2 }{\mathcal{H}\left( { \sqrt{q_0}z_{0}+\sqrt{q_1-q_0}z_{1}}, {\sqrt{1-q_1}} \right) }\,,
\end{equation}

where $G(x)$ denotes the standard Gaussian density function.
Substituting \eqref{eq:qh1(q1)} in \eqref{eq:fq1a}, we have $\frac{\partial \tilde \phi}{\partial \hat q_1}$ as a function of $q_1$ only. We want to show that for any $\beta > 0$, this derivative vanishes in the limit  $q_1 = 1$. To do so, take
\begin{equation}
    \label{app:q1_to_1}
    q_1 = 1-\epsilon \,.
\end{equation}
Then, we compute $\hat q_1$ from \eqref{eq:qh1(q1)}, as a function of $\epsilon \simeq 0$:
\begin{align*}
    \hat q_1 & \simeq \alpha \frac{(1-e^{-  \beta})^2}{\epsilon}
    \int Dz_0 \, \frac{2H\left(-\frac{r z_0}{\sqrt{q_0-r^2}}\right)  } {\mathcal{H}\left( {\sqrt{q_0}z_{0} }, {\sqrt{1-q_0}}\right)}
    \int Dz_1 \frac{  G\left( \frac{\sqrt{q_0}z_{0}+\sqrt{1-q_0}z_{1}}{\sqrt{\epsilon}} \right) ^2 }{ \mathcal{H}\left( { \sqrt{q_0}z_{0}+\sqrt{1-q_0}z_{1}},{\sqrt{\epsilon}} \right) } \\
             & \simeq \alpha \frac{(1-e^{- \beta})^2}{\epsilon}
    \int Dz_0 \, \frac{ 2H\left(-\frac{r z_0}{\sqrt{q_0-r^2}}\right) } {\mathcal{H}\left(  {\sqrt{q_0}z_{0} },{\sqrt{1-q_0}}\right)}
    \frac{\sqrt{\epsilon}}{\sqrt{1-q_0}} G\left(\frac{\sqrt{q_0}z_{0} }{\sqrt{1-q_0}}\right) \int du \frac{  G\left( u \right) ^2 }{ \mathcal{H}\left( u, 1 \right)}                     \\
             & = \frac{C(q_0, r, \alpha, \beta)}{\sqrt{\epsilon}}\,,
\end{align*}
where
\begin{equation}
    C(q_0, r, \alpha, \beta) = \alpha\, \frac{(1-e^{-\beta})^2}{\sqrt{1-q_0}}
    \int Dz_0 \, \frac{2H\left(-\frac{r z_0}{\sqrt{q_0 - r^2}}\right)  G\left(\frac{\sqrt{q_0}z_{0} }{\sqrt{1-q_0}}\right)} {\mathcal{H}\left({\sqrt{q_0}z_{0} },{\sqrt{1-q_0}} \right)} \, \int du \frac{  G\left( u \right) ^2 }{ \mathcal{H}\left( u, 1 \right)}\,.
\end{equation}
Notice that $C(q_0, r, \alpha, \beta)>0$ when $\beta >0$ and 0 if $\beta = 0$. For small $\beta$, we can use $\mathcal{H}(x, y)^{-1} \simeq 1 + O(\beta)$
, so that
\begin{equation}
    C(q_0, r, \alpha, \beta) = c(\alpha, q_0)\,\beta^2 + O\left(\beta^3\right) \,.
\end{equation}
Then, we consider the saddle point equation~\eqref{eq:fq1a}. Rewrite it as a self-consistent equation:
\begin{align}
    \label{eq:spe_abp_td}
    q_1 = f(\hat{q}_1(q_1))\,,
\end{align}
where $f(\hat{q}_1)$ is given by the right-hand-side of \eqref{eq:fq1a}:
\begin{align*}
    f(\hat{q}_1) 
     & = e^{-\frac{\hat q_1 - \hat q_0}{2}} \int Dz_0  Dz_1  \frac{\cosh \left(\sqrt{\hat q_0} z_0 + \sqrt{\hat q_1 - \hat q_0} z_1 +\hat{r}\right) - [\cosh\left(\sqrt{\hat q_0} z_0 + \sqrt{\hat q_1 - \hat q_0} z_1 +\hat{r} \right)]^{-1}}{\cosh (\sqrt{\hat q_0} z_0+\hat{r})}
\end{align*}
and $\hat{q}_1(q_1)$ is given by \eqref{eq:qh1(q1)}. We want to show that $f(\hat{q}_1(q_1)) \to 1$ in the limit $q_1 \to 1$, satisfying a consistency relation. We have:
\begin{equation}
    \begin{split}
        \label{eq::scaling_f}
        f\left(\frac{C}{\sqrt{\epsilon}}\right) & \simeq e^{-\frac{\hat q_1 - \hat q_0}{2}} \int \frac{Du}{\cosh (\sqrt{\hat q_0} u + \hat{r})} \, \left[ e^{\frac{\hat q_1 - \hat q_0}{2}}\cosh (\sqrt{\hat q_0} u + \hat{r}) - \sqrt{\frac{\pi}{2}} \frac{{\epsilon}^{1/4}}{C^{1/2}} + o\left(\frac{\epsilon^{3/4}}{C^{3/2}}\right) \right] \\
                                                & \simeq 1 - \sqrt{\frac{\pi}{2}} e^{-\frac{C}{2\sqrt{\epsilon}}} \frac{\epsilon^{1/4}}{C^{1/2}}\,,
    \end{split}
\end{equation}
having used $\int_{-\infty}^{\infty} dx \, \cosh(x)^{-1} = \pi$. Since for $\epsilon \to 0$ one has $f \to 1$ when $C>0$, we have shown that the frozen solution $q_1 = 1$ is always a solution for any $\beta > 0$. So $\beta_d = 0$ in the thermodynamic limit.
Finally, one can also compute the behavior of $\beta_d$ as $\epsilon\to 0$, by solving \eqref{eq:spe_abp_td} for small $\beta$:
$$
    1-\epsilon = 1 - \sqrt{\frac{\pi}{2}} e^{-\frac{c\beta^2}{2\sqrt{\epsilon}}} \frac{\epsilon^{1/4}}{c^{1/2}\beta}
$$
In the small $\epsilon$ limit, $\beta_d$ vanishes as:
\begin{equation}
    \beta_d \sim \epsilon^{1/4} \sqrt{\ln\left(\frac{1}{\epsilon}\right)}\,.
\end{equation}
To summarize, our computation reveals a divergent dynamical temperature in the large $N$ limit:
$$\lim_{N\to\infty} T_d = +\infty \,. $$
In order to find a possible scaling of $T_d$ with $N$, we
set the correction $\epsilon$ appearing in \eqref{app:q1_to_1} to:
$$ \epsilon = 1- q_1 \simeq \frac{1}{N} \,.$$
This heuristic scaling comes directly from the definition of the Hamming distance $d_H$ between two binary vectors. When $N$ is finite, the size of the point-like clusters is rounded by the discreteness of the configuration space. The smallest possible Hamming distance between two binary vectors occurs when they differ by a single bit; this corresponds to an Hamming distance equal to $d_H = 2 / N$.
This argument allows one to identify an heuristic scaling of $\beta_d$ with $N$:
\begin{equation}
    \label{appx:scaling}
    \beta_d \sim \frac{\sqrt{\ln\left(N\right)}}{N^{1/4}} \,.
\end{equation}
We point out that this result does not depend on $r$ and $\hat{r}$, and that the exact same scaling is found for the SBP, where we have a further simplification due to the fact that $q_0 = \hat{q}_0 = 0$.

\subsection{General criterion for freezing} \label{app::generic_criterion_freezing}
The previous computation for the ABP shows that the existence of the frozen solution is controlled by the behaviour of the energetic saddle-point equation \eqref{eq:fqh1a}:
\begin{equation}
    \begin{split}
        \hat q_1 & =
        \alpha\int Dz_0\,
        \frac{
            2H\left(-\frac{r z_0}{\sqrt{q_0-r^2}}\right)
        }{
            \mathcal H\left(\sqrt{q_0}z_0,\sqrt{1-q_0}\right)
        }
        \int Dz_1\,
        \frac{
            \left[
                \partial_x \mathcal H
                \left(
                \sqrt{q_0}z_0+\sqrt{q_1-q_0}z_1,
                \sqrt{1-q_1}
                \right)
                \right]^2
        }{
            \mathcal H
            \left(
            \sqrt{q_0}z_0+\sqrt{q_1-q_0}z_1,
            \sqrt{1-q_1}
            \right)
        } \\
                 & = \alpha\int Dz_0\,
        \frac{
            2H\left(-\frac{r z_0}{\sqrt{q_0-r^2}}\right)
        }{
            \mathcal H\left(\sqrt{q_0}z_0,\sqrt{1-q_0}\right)
        }
        \int \frac{dx}{\sqrt{q_1-q_0}}\, G\left(\frac{x - \sqrt{q_0} z_0 }{\sqrt{q_1-q_0}} \right)
        \frac{
            \left[
                \partial_x \mathcal H
                \left( x,
                \sqrt{1-q_1}
                \right)
                \right]^2
        }{
            \mathcal H
            \left(
            x,
            \sqrt{1-q_1}
            \right)
        }
    \end{split}
    \label{eq:general_qhat1_boundary}
\end{equation}
The entropic saddle-point equation, instead, is independent of the particular energetic factor. It can be written as
\begin{equation}
    q_1 = f(\hat q_1),
\end{equation}
and, for $\hat q_1\to\infty$, one has as shown in~\eqref{eq::scaling_f}
\begin{equation}
    f(\hat q_1)
    =
    1
    -
    \sqrt{\frac{\pi}{2}} \,
    \hat q_1^{-1/2}
    e^{-\hat q_1/2}
    \left[1+o(1)\right],
    \label{eq:f_large_qhat1}
\end{equation}
Therefore the frozen solution $q_1=1$ can be obtained only if the energetic equation sends $\hat q_1$ to infinity when $q_1\to1$. The singularity of \eqref{eq:general_qhat1_boundary} is entirely determined by the small-$y$ behaviour of
\begin{equation}
    \frac{\left[\partial_x \mathcal H(x,y)\right]^2}{\mathcal H(x,y)},
    \qquad
    y=\sqrt{1-q_1} 
\end{equation}
In order to make this statement more explicit, let us go back to the
definition of the kernel
\begin{equation}
    \mathcal H(x,y)=\int Dz\,\mathcal K(x+yz),
    \label{eq:H_from_K_boundary}
\end{equation}
where $\mathcal K$ is the single-pattern Boltzmann factor, Cf. Eqs. \eqref{eq:K_ABP_main},  \eqref{eq:K_SBP_main}. Thus $\mathcal H(x,y)$ is a Gaussian smoothing of $\mathcal K$ on a
scale $y$. If $\mathcal K$ is smooth around a point $x$, then
$\partial_x\mathcal H(x,y)$ remains finite as $y\to0$. Singular
contributions can only arise close to points where $\mathcal K$ is
non-smooth. This usually happens near the decision boundaries of the constraints.

Let $b$ be such a decision boundary point. In the $x$ integral appearing in
\eqref{eq:general_qhat1_boundary}, the Gaussian density multiplying
$(\partial_x\mathcal H)^2/\mathcal H$ is smooth on the scale $y$.
Therefore, close to $b$, one can set
\begin{equation}
    x=b+yu,
    \qquad
    dx=y\,du.
    \label{eq:boundary_layer_variable}
\end{equation}
The whole question is then
reduced to the scaling with $y$ of the \textit{boundary layer} term:
\begin{equation}
    \mathcal B_{\mathcal K}(b,y) \equiv y\int du\,
    \frac{
        \left[\partial_x\mathcal H(b+yu,y)\right]^2
    }{
        \mathcal H(b+yu,y)
    } .
    \label{eq:boundary_layer_integral}
\end{equation}
If this quantity diverges as $y\to0$, then
$\hat q_1\to\infty$. Since the
entropic equation satisfies \eqref{eq:f_large_qhat1}, this implies that
the frozen solution $q_1=1$ is present. If instead
\eqref{eq:boundary_layer_integral} stays finite, or vanishes, the
energetic equation does not force $\hat q_1\to\infty$, and the frozen solution at $q_1=1$ is absent.

We now discuss the possible behaviors of $\mathcal K$ at a boundary.

\begin{itemize}
    \item \emph{Jump discontinuity.}
          Suppose that, close to $b$, the Boltzmann factor has a finite jump:
          \begin{equation}
              \mathcal K(s)
              =
              \mathcal K_-
              +
              \Delta\mathcal K\,\Theta(s-b),
              \qquad
              \Delta\mathcal K\neq0.
              \label{eq:K_jump}
          \end{equation}
          This is the case for both \eqref{eq:K_ABP_main} and \eqref{eq:K_SBP_main} for $b=0$ and $b=\pm \kappa$ respectively.
          Then, using \eqref{eq:H_from_K_boundary} and setting $x=b+yu$,
          \begin{equation}
              \mathcal H(b+yu,y)
              =
              \mathcal K_-
              +
              \Delta\mathcal K
              \int Dz\,\Theta(u+z).
          \end{equation}
          Thus $\mathcal H(b+yu,y)$ is of order one inside the boundary layer.
          On the other hand,
          \begin{equation}
              \partial_x\mathcal H(b+yu,y)
              =
              \Delta\mathcal K
              \int Dz\,\delta(b+yu+yz-b)
              =
              \frac{\Delta\mathcal K}{y}\,G(u).
          \end{equation}
          Therefore
          \begin{equation}
              \mathcal B_{\mathcal K}(b,y)
              \simeq
              \frac{(\Delta\mathcal K)^2}{y}
              \int du\,
              \frac{
                  G(u)^2
              }{
                  \mathcal K_-+\Delta\mathcal K\int Dz\,\Theta(u+z)
              } .
          \end{equation}
          Hence a jump produces the divergence
          \begin{equation}
              \mathcal B_{\mathcal K}(b,y)
              = O\left(\frac{1}{y}\right) = O\left(\frac{1}{\sqrt{1-q_1}}\right)
              \label{eq:jump_boundary_scaling}
          \end{equation}
          This is precisely the mechanism found in the ABP/SBP computation done in the previous subsection.  Indeed for the ABP: $\mathcal{K}_{-} = e^{-\beta}$, $\Delta\mathcal K=1-e^{-\beta}$ and $b=0$.

    \item \emph{Vanishing power at the boundary.}
          Suppose instead that the Boltzmann factor vanishes continuously at the boundary as
          \begin{equation}
              \mathcal K(s)
              \simeq
              A(s-b)^\gamma\Theta(s-b),
              \qquad
              A, \gamma >0.
              \label{eq:K_power_vanishing}
          \end{equation}
          see the right panel of Figure~\ref{fig:single_pattern_weights}. Then
          \begin{align}
              \mathcal H(b+yu,y) & = \int Dz\,\mathcal K(b+y(u+z)) \simeq A y^\gamma \int Dz\,(u+z)^\gamma \Theta(u+z) \equiv A y^\gamma F_\gamma(u)
          \end{align}
          where $F_\gamma(u)=\int Dz\,(u+z)^\gamma\Theta(u+z)$.
          We therefore have
          \begin{equation}
              \partial_x\mathcal H(b+yu,y)
              \simeq A y^{\gamma-1}F_\gamma'(u).
          \end{equation}
          Therefore
          \begin{equation}
              \mathcal B_{\mathcal K}(b,y)
              \simeq
              y^{\gamma-1}
              \int du\,\frac{F_\gamma'(u)^2}{F_\gamma(u)}.
              \label{eq:power_boundary_scaling}
          \end{equation}
          Consequently, if $\gamma<1$, $\hat q_1$ diverges and one has the frozen solution $q_1=1$, whereas if $\gamma>1$ the frozen solution is lost. The case $\gamma=1$ is marginal. This criterion also explains why the logarithmic potential studied in~\cite{straziota2026generative}, which falls in this category, can remove freezing.

    \item \emph{Continuous non-zero value at the boundary.}
          Another possibility for the behavior of the kernel near the boundary point is the following
          \begin{equation}
              \mathcal K(s) = \mathcal K_0
              + A(b-s)^\gamma\Theta(b-s) \,,
              \qquad \mathcal K_0>0.
              \label{eq:K_continuous_nonzero}
          \end{equation}
          In this case
          \begin{equation}
              \mathcal H(b+yu,y) = \mathcal K_0 + O(y^\gamma),
          \end{equation}
          while
          \begin{equation}
              \partial_x\mathcal H(b+yu,y) = O(y^{\gamma-1}).
          \end{equation}
          Hence
          \begin{equation}
              \mathcal B_{\mathcal K}(b,y) \sim y^{2\gamma-1}.
              \label{eq:continuous_boundary_scaling}
          \end{equation}
          Therefore, for ordinary integer powers $\gamma>1/2$, there is no divergent boundary contribution. This category includes the single pattern Gibbs weight
          \begin{equation}
              \mathcal K(s) = e^{-\beta(-s)^\gamma \Theta(-s)} = e^{-\beta(-s)^\gamma} + \left(1-e^{-\beta(-s)^\gamma} \right) \Theta(s)
          \end{equation}
          studied by Horner~\cite{Horner-storage} and by~\cite{copycat}, who focused on positive integer values of $\gamma$; see also the left panel of Figure~\ref{fig:single_pattern_weights} for a plot. At finite $\beta$ this factor is continuous at $s=0$:
          \begin{equation}
              \mathcal K(0^-)=\mathcal K(0^+)=1.
          \end{equation}
          Close to the boundary,
          \begin{equation}
              \mathcal K(s) = 1-\beta(-s)^\gamma\Theta(-s)+\cdots.
          \end{equation}
          This is of the form \eqref{eq:K_continuous_nonzero} with $\mathcal{K}_0 = 1$, $A = -\beta$.
          For the usual integer cases $\gamma\ge1$, the boundary contribution does not diverge. Hence,
          at finite $\beta$, the energetic saddle equation does not force $\hat q_1\to\infty$ as $q_1\to1$, and the frozen solution
          disappears. A dynamical transition may still occur, but at a finite temperature and with $q_1<1$ at the transition. At zero temperature, however, as $\mathcal K(s)\to \Theta(s)$, the jump is restored, and the freezing mechanism reappears.

\end{itemize}

\section{Detail of the message passing algorithms}
\label{app:ramp_algorithm}

In this section, we provide details of the message passing algorithm used in the main text to search for minimal error configurations in the ABP. We start from a short review of the AMP algorithm, then we move to the description of AMP+reinforcement.

\subsection{Approximate Message Passing}
\begin{figure}[t]
    \centering
    \includegraphics[width=0.33\linewidth]{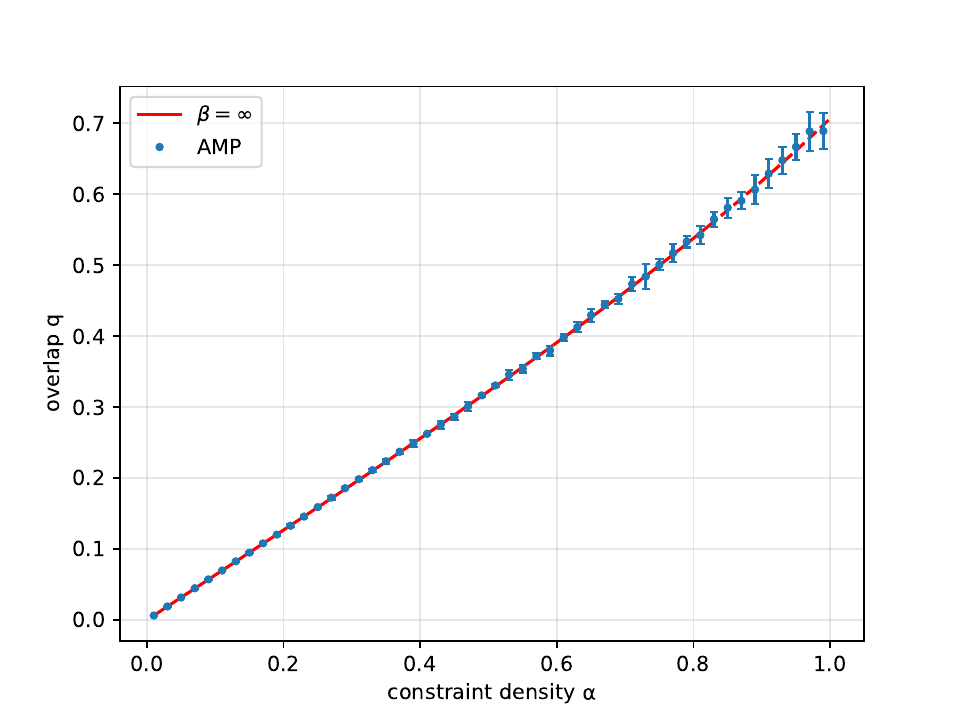}\hfill
    \includegraphics[width=0.33\linewidth]{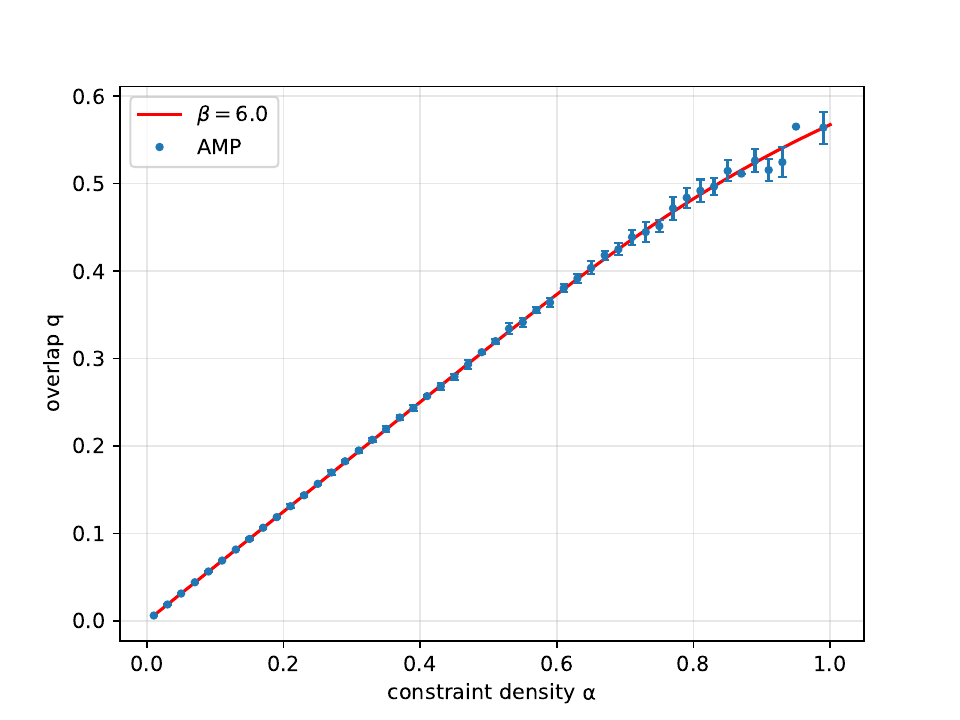}\hfill
    \includegraphics[width=0.33\linewidth]{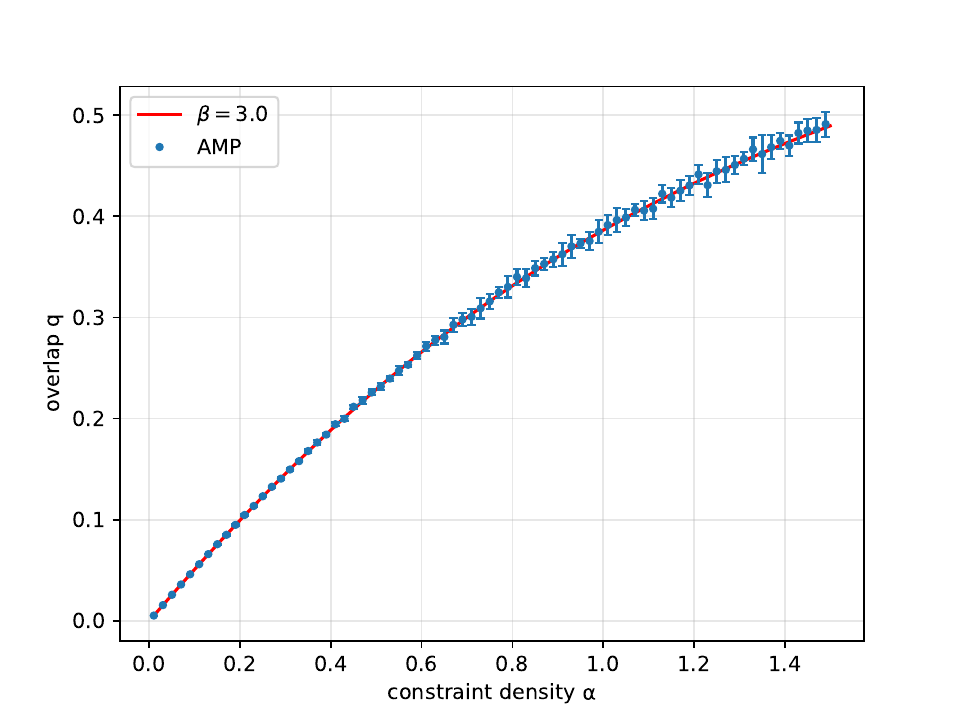}
    \caption{The AMP algorithm was tested at different values of temperature, corresponding to $\beta=\infty$, $\beta=6.0$, $\beta=3.0$ (left to right), in the storage setting. The plots present the typical RS overlap $q$ as a function of the constraint density $\alpha$: experimental points (blue) were obtained with $N=1000$ and averaged over $20$ samples, and they agree with the theoretical RS curve (red).}
    \label{fig:AMP}
\end{figure}

Consider the Gibbs measure induced by the partition function~\eqref{eq:generic_partition_main}:
\begin{align}\label{eq:posterior}
    p_{\mathcal{D}} \left(\bm w\right) = \frac{\prod_{\mu=1}^{P} \mathcal{K} \left( s^\mu(\bm{w}) \right) }{Z_{\mathcal{D}}}\,. 
\end{align}
The Approximate Message Passing (AMP) algorithm provides an iterative approximation to the local marginals of the measure~\eqref{eq:posterior}.
AMP is obtained from the Belief Propagation (BP) equations using a Gaussian approximation which parametrizes the one site marginals $m_i(w_i) = \sum_{\boldsymbol{w}_{\backslash i}} \, p_{\mathcal{D}}(\boldsymbol{w})$ in terms of its mean $a_i$ and variance $b_i$.  Being the weights binary in our setting, the one site mean
\begin{align}
    a_i & = \sum_{ w_i = \pm 1 } \, m_i(w_i)\, w_i \equiv \langle w_i \rangle_{\beta} 
\end{align}
completely determines also the variance of the one site marginal via $b_i = 1 - a_i^2$. Starting at $t=0$ from a random guess for the one site marginals $a_i^{t=0}$ and initializing randomly the $P$ dimensional vector $g_\mu^{t=0}$ the AMP algorithm consists in the following update equations:
\begin{subequations}
    \begin{align}
        V_\mu^t & = \sum_{i=1}^N \frac{(x_i^\mu)^2}{N} \left[1-\left(a_i^{t-1}\right)^2\right]\,, 
        \\
        M_\mu^t & = \sum_{i=1}^N \frac{x_i^\mu}{\sqrt N} a_i^{t-1} - V_\mu^t g_\mu^{t-1}\,,       
        \\
        g_\mu^t & = g_E(y^\mu,M_\mu^t,V_\mu^t;\beta)\,,                                           \\
        \label{eq::AMP_local_field}
        h_i^t   & = \sum_{\mu=1}^P \frac{x_i^\mu}{\sqrt N} g_E(y^\mu,M_\mu^t,V_\mu^t;\beta)
        - a_i^{t-1} \sum_{\mu=1}^P \frac{(x_i^\mu)^2}{N} \partial_M g_E(y^\mu,M_\mu^t,V_\mu^t;\beta)\,,
        \\
        a_i^t   & = \tanh(h_i^t)\,,
    \end{align}
\end{subequations}
which are iterated for $t=1, \dots, t_{\rm max}$ or until convergence. The function $g_E$ is called the \emph{energetic channel} and depends in general on the detail of the model. For example in the case of ABP with error counting loss function it reads:
\begin{equation}
    g_E(y,M,V;\beta) = \partial_M \log\left[ e^{-\beta} + \left(1-e^{-\beta}\right) H\left(-\frac{yM}{\sqrt V}\right) \right],
    \label{eq:algorithmic_gE}
\end{equation}
where we recall that $H(x) \equiv \frac{1}{2} \mathrm{Erfc}\left(\frac{x}{\sqrt{2}}\right)$.  When AMP converges, it can be used to compute on a single sample other observables that are computed via the replica method. For example the overlap between two students sampled from \eqref{eq:posterior} can be written in terms of the one-site magnetizations as:
\begin{equation}
    q_N = \frac{1}{N}\sum_{i=1}^N \langle w_i \rangle_{\beta} \langle w_i \rangle_{\beta} = \frac{1}{N}\sum_{i=1}^N a_i^2\,.
\end{equation}
In the large $N$ limit, this can be proved to converge to the replica method RS order parameter $q$~\cite{MM-book}.
As a check of the reliability of our algorithm at finite temperature we confirmed this by running a few experiments, as reported in Figure \ref{fig:AMP}.

\subsection{AMP + reinforcement (rAMP)}

The reinforcement AMP algorithm (rAMP)~\cite{Braunstein_2006} is a modification of the AMP iteration designed to turn the soft marginal information
computed by AMP into an actual binary configuration. Standard AMP estimates the local magnetizations $a_i\simeq \langle w_i\rangle$ of the Gibbs measure, but these magnetizations need not be close to $\pm1$. The role of reinforcement is to progressively polarize the local fields, so that the magnetizations become closer to the vertices of the hypercube and the configuration given by
\begin{equation}
    w_i=\sign(a_i)
\end{equation}
can be used as a candidate low-error assignment. Operationally, reinforcement is implemented by adding to the AMP field
update a term proportional to the previous local field,
\begin{equation}
    h_i^t
    =
    h_{i,\mathrm{AMP}}^t
    +
    \rho_{t-1} h_i^{t-1},
\end{equation}
where $h_{i,\mathrm{AMP}}^t$ denotes the standard AMP update, see~\eqref{eq::AMP_local_field}. The parameter $\rho_{t}$ is increased during the dynamics according to a reinforcement schedule. In the experiments reported in the main text we use
\begin{equation}
    \rho_t = 1-(1-\rho)^t ,
    \label{eq:reinforcement_schedule_appendix}
\end{equation}
where $\rho$ controls the speed at which the reinforcement is switched on. For small $\rho$ and early times, $\rho_t\simeq \rho t$, while at longer times the reinforcement strength approaches one. Setting $\rho=0$ gives $\rho_t=0$ for all $t$, and one recovers the standard AMP iteration.

\begin{algorithm}[H]
    \caption{One AMP+reinforcement update}
    \label{alg:ramp_step}
    \begin{algorithmic}

        \Function{rAMPstep}{$\bm h^{t-1},\bm a^{t-1},\bm g^{t-1};\rho_{t-1},\beta$}

        \State Compute the variances
        \Statex
        \begin{equation}
            V_\mu^t
            \leftarrow
            \sum_{i=1}^N
            \frac{(x_i^\mu)^2}{N}
            \left[1-\left(a_i^{t-1}\right)^2\right],
            \qquad
            \mu=1,\ldots,P .
        \end{equation}

        \State Compute the Onsager-corrected means
        \Statex
        \begin{equation}
            M_\mu^t
            \leftarrow
            \sum_{i=1}^N
            \frac{x_i^\mu}{\sqrt N}
            a_i^{t-1}
            -
            V_\mu^t g_\mu^{t-1},
            \qquad
            \mu=1,\ldots,P .
        \end{equation}

        \State Update the energetic messages
        \Statex
        \begin{equation}
            g_\mu^t
            \leftarrow
            g_E(y^\mu,M_\mu^t,V_\mu^t;\beta),
            \qquad
            \mu=1,\ldots,P .
        \end{equation}

        \State Update the fields
        \Statex
        \begin{align}
            h_i^t
            \leftarrow
             &
            \sum_{\mu=1}^P
            \frac{x_i^\mu}{\sqrt N}
            g_\mu^t
            -
            a_i^{t-1}
            \sum_{\mu=1}^P
            \frac{(x_i^\mu)^2}{N}
            \partial_M g_E(y^\mu,M_\mu^t,V_\mu^t;\beta) + \rho_{t-1}h_i^{t-1},
            \qquad
            i=1,\ldots,N .
        \end{align}

        \State Update the magnetizations
        \Statex
        \begin{equation}
            a_i^t
            \leftarrow
            \tanh(h_i^t),
            \qquad
            i=1,\ldots,N .
        \end{equation}

        \State \Return $(\bm h^t,\bm a^t,\bm g^t)$.

        \EndFunction

    \end{algorithmic}
\end{algorithm}

\begin{figure}[b]
    \centering
    \includegraphics[width=0.5\linewidth]{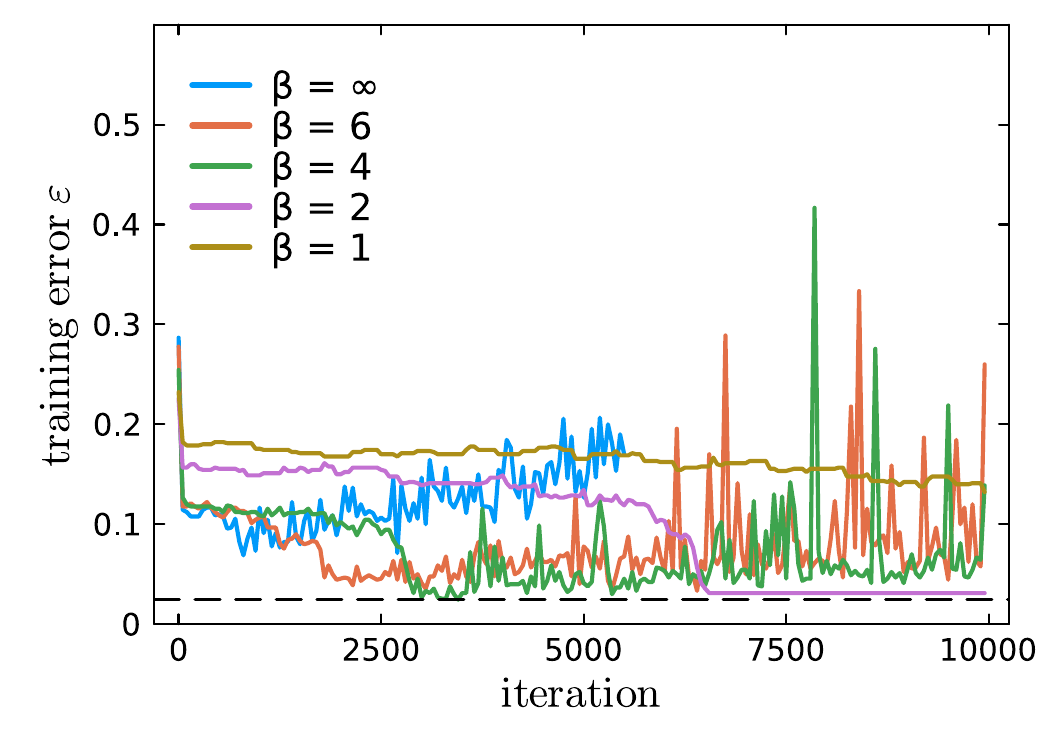}\hfill
    \caption{ABP, storage setting. Training-error trajectories of rAMP for $N=1000$, $\alpha=0.9$, reinforcement rate $\rho=10^{-4}$, and different inverse temperatures $\beta$. All curves use the same dataset and initialization. The dashed horizontal line marks the target training error $\epsilon^{\mathrm{targ}}=0.025$. Among the values considered, only $\beta=4$ reaches the target within $10^{4}$ iterations.}
    \label{fig:rAMP}
\end{figure}

The explicit one-step rAMP update is given in Algorithm~\ref{alg:ramp_step}. The fixed-temperature reinforced dynamics, obtained by iterating this update at fixed $\beta$, is summarized in Algorithm~\ref{alg:ramp_fixed_beta} in the main text.

We show in Figure~\ref{fig:rAMP} the behavior of the rAMP algorithm for fixed reinforcement rate and several values of $\beta$ for $\alpha = 0.9$ and $N=1000$ in the storage setting of the ABP. Notice that the $\beta=\infty$ trajectory stops before reaching the maximum number of iterations, because the rAMP produces diverging updates. This is expected as for $\alpha>0.784$ it is hard to find solutions, and in addition for $\alpha$ greater than the SAT/UNSAT threshold $\alpha_S = 0.833$ solutions do not exist at all in the large $N$ limit \cite{Krauth}. Decreasing the value of $\beta$ allows to search for configurations having a certain positive value of the training error. Moreover, the resulting AMP updates are less singular and the local marginals are initially less polarized, which makes the subsequent reinforcement dynamics more stable. If one continues decreasing $\beta$ but maintaining fixed the reinforcement rate $\rho$, the trajectory tends to stabilize at a higher training error value before exploring lower training error configurations at a larger number of iterations. The figure therefore shows that for each fixed reinforcement rate, there exists an optimal value of $\beta$ that allows to probe the lowest achievable training error configurations.

We also tested a temperature-annealing schedule. However, this introduces some additional hyperparameters: the starting temperature $\beta_0$, the temperature increment or cooling rate $\Delta \beta$ and the frequency of increment $\Delta_{\rm it}$. Those new hyperparameters need to be tuned optimally together with the reinforcement schedule. Overall, we found no substantial improvement from incorporating temperature annealing.

\section{The generalization error}\label{app:sec_gen_err}
Let us consider again the perceptron problem studied at equilibrium, in the teacher-student setting. We recall that the generalization error committed by a typical student is:
\begin{equation}\label{appx:eg}
    \epsilon_g = \mathbb{E}_{\tilde{\bm{x}}} \mathbb{E}_{\mathcal D}\mathbb{E}_{\bm{w}|\mathcal{D}}\, \left[\Theta
        (-y\, \hat{y}) \right]= \frac{1}{\pi}\arccos(r)\,,
\end{equation}
where $\tilde{\bm{x}}$ represents a test pattern, following the same distribution as the training data $\mathcal{D}$ and whose label $y$ is assigned by the teacher $\bm w^\star$, $\mathbb{E}_{\bm w | \mathcal D}[\bullet]$ represents the average over the Gibbs measure \eqref{eq:generic_partition_main}, and finally
$$
    \hat y= \sign\left( \frac{\bm{w}\cdot\tilde{\bm{x}}}{\sqrt{N}}\right)
$$
represents the student output. Expression \eqref{appx:eg} can also be used to compute the generalization error committed by the typical student $\bm w$ in a constrained $m$-tuple, by averaging over the measure \eqref{eq::cloned_partition}. Notice that the equilibrium value of the typical teacher-student overlap $r$ in \eqref{appx:eg} is determined by the saddle point equations, and therefore on the external parameters $\alpha$, $\beta$, and $q_1$ and $m$ in the cloned system.

In an inference framework, one aims at building an estimator $\hat{\bm{w}}$ with good generalization performance.
Here, we consider the \textit{barycenter} of typical students, trained on the dataset $\mathcal{D}$ and weighted according to the Gibbs measure at inverse temperature $\beta$, namely:
\begin{equation}
    \hat{\bm{w}} = \mathbb{E}_{\bm w|\mathcal{D}}[\bm{w}]\,.
\end{equation}
This allows one to compute the so-called \textit{Bayesian} generalization error, representing
the probability that the ensemble of students trained on $\mathcal{D}$ fails the classification of test data points.
In formulas:
\begin{equation}\label{app:Bayes_error}
    \epsilon_g^{\rm Bayes} = \mathbb{E}_{\tilde{\bm x}}\mathbb{E}_{\mathcal D}\, \Theta \left( - {y} \; \mathbb{E}_{\bm{w}|\mathcal D}\left[ \hat{y}
        \right]\right).
\end{equation}
A replica calculation \cite{Learning_through_atypical} provides one with a closed expression for the Bayesian error, which depends on both the teacher-student overlap $r$ and on the students overlap $q$:
\begin{equation}\label{}
    \epsilon_g^{\rm Bayes}
    =
    \frac{1}{\pi}\arccos \left(\frac{r}{\sqrt{q}}\right).
\end{equation}
The Bayesian error is therefore guaranteed to be smaller than the error of the single student $\epsilon_g$, at any temperature $\beta$ and sample density $\alpha$. 

\subsection{Additional figures}\label{appx:additional_figures}

\begin{figure}[H]
    \centering
    \includegraphics[width=0.5\linewidth]{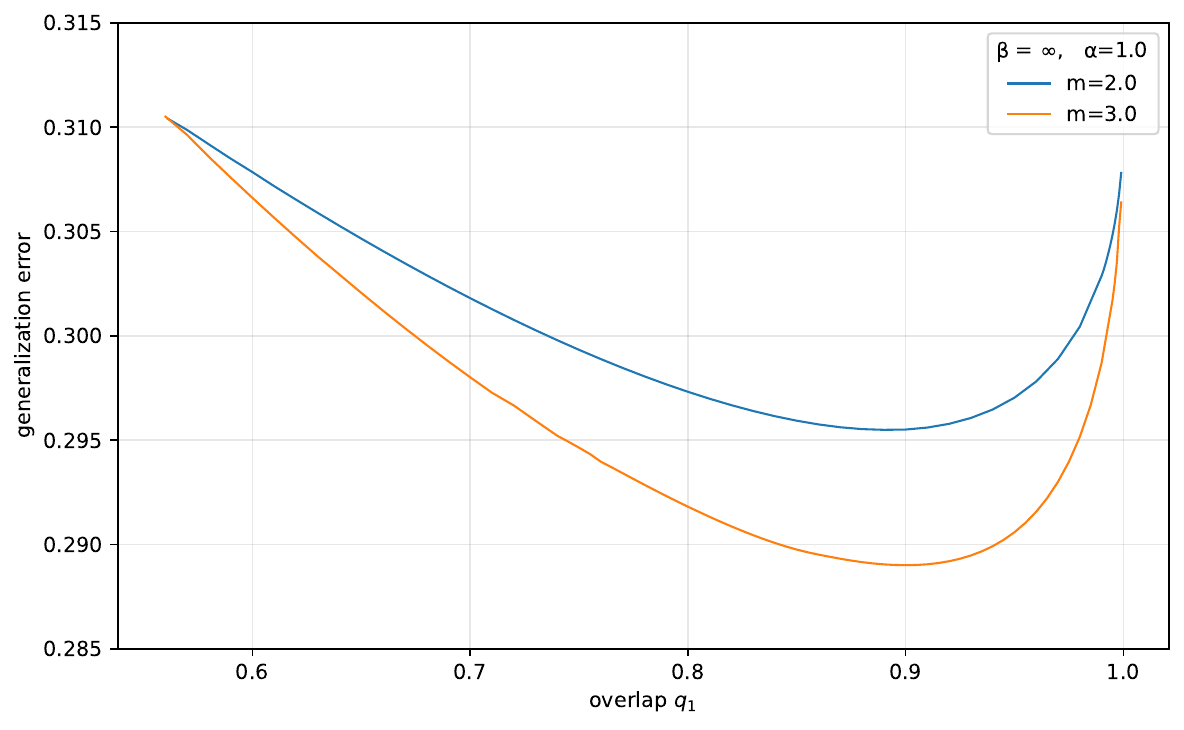}\hfill
    \includegraphics[width=0.5\linewidth]{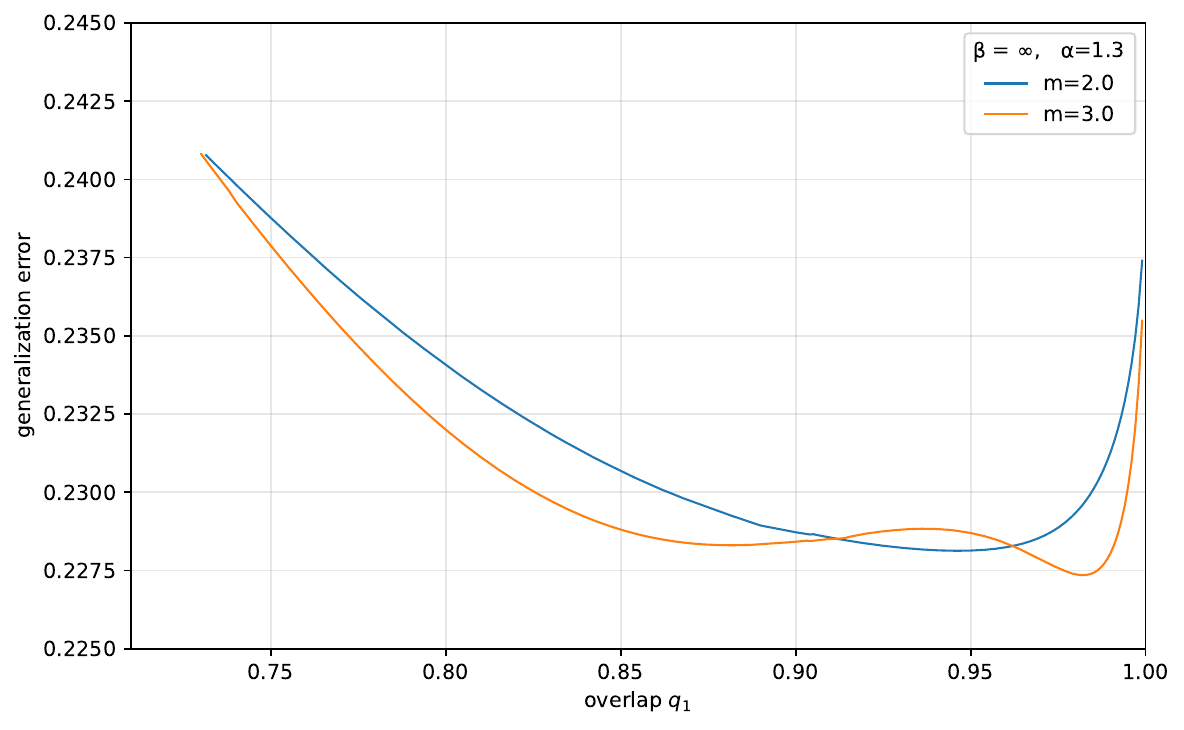}
    \caption{Generalization error as a function of the overlap $q_1$ between the clones, for $m=2$ and $m=3$, and $\beta=\infty$. The left picture corresponds to the easy phase of Figure \ref{fig:generalization} ($\alpha=1.0$), while the right figure to the hard phase ($\alpha=1.3$) and all its points are characterized by a negative entropy. Clusters of $m=3$ clones can achieve a better generalization, if the hyperparameter $q_1$ is tuned conveniently.}
    \label{fig:placeholder}
\end{figure}

\begin{figure}[H]
    \centering
    \includegraphics[width=0.5\linewidth]{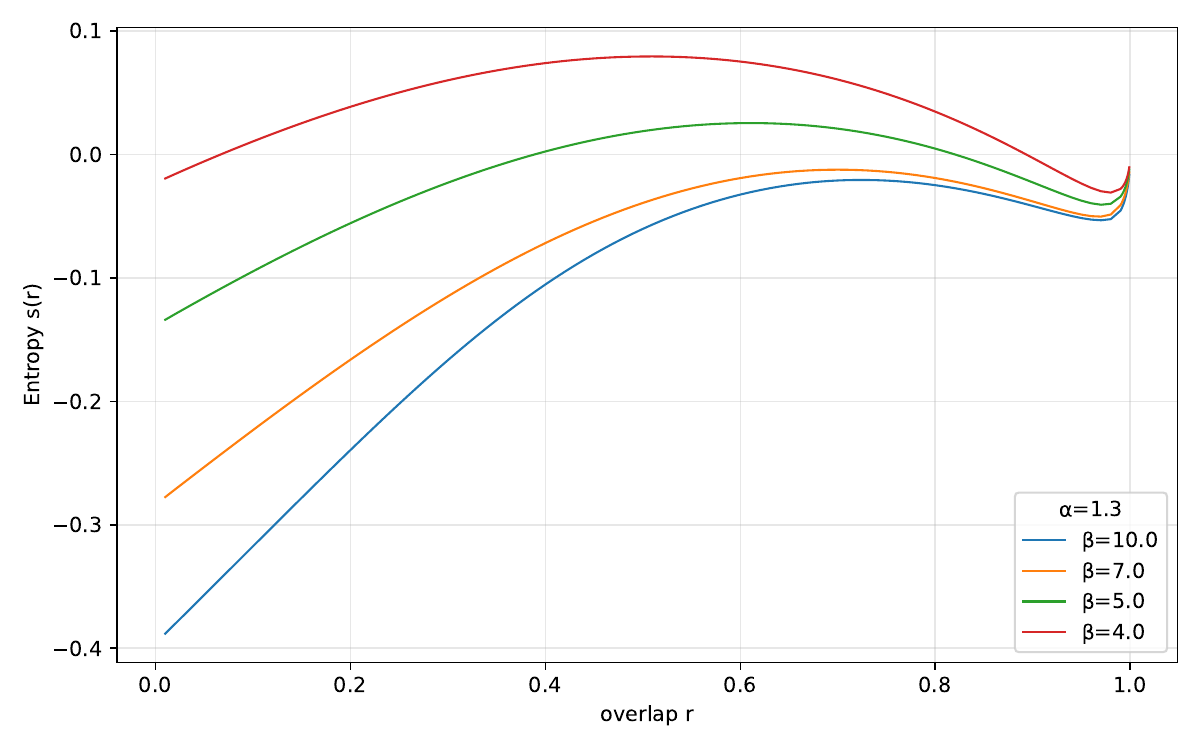}\hfill
    \includegraphics[width=0.5\linewidth]{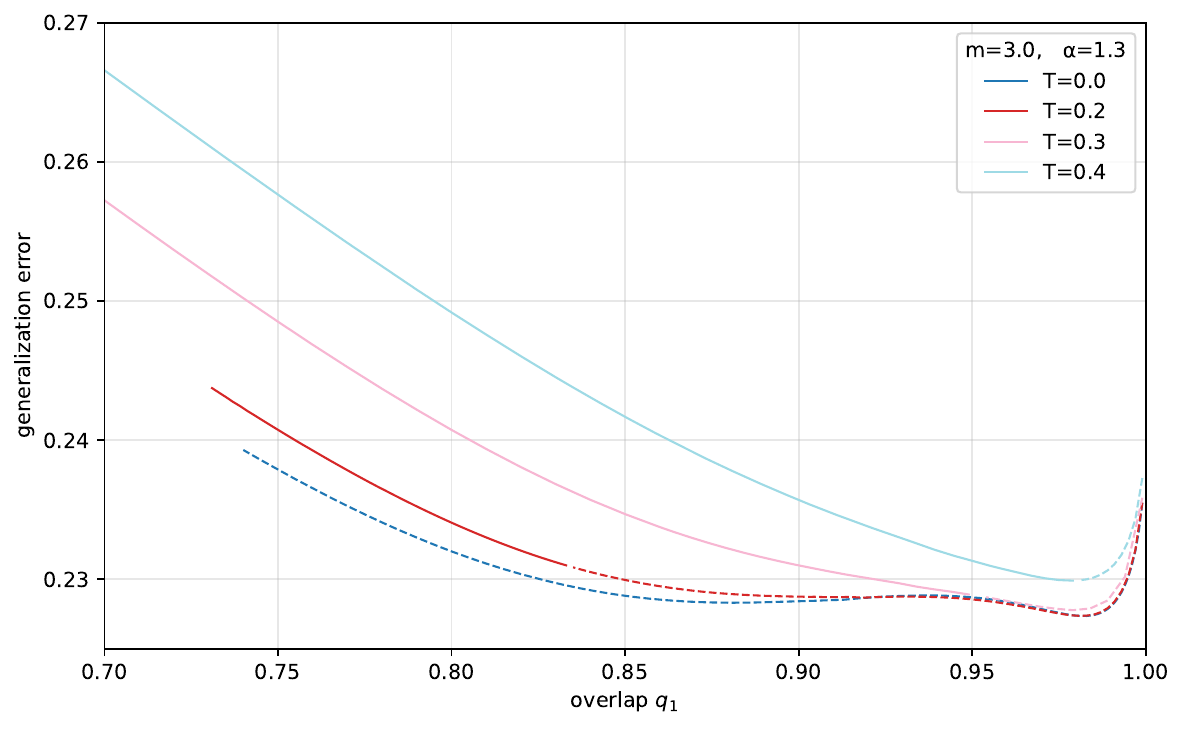}
    \caption{Left: the picture shows the entropy of the RS solution, as a function of the teacher-student overlap $r(\bm{w}, \bm{w}^\star)$, in the hard phase ($\alpha=1.3$). The entropy increases with temperature, as expected. The same happens with solution of the $3$-cloned system: the right picture shows the generalization error of a solution in the cluster, as a function of the mutual overlap $q_1$, similarly to Figure \ref{fig:TS_generalization_anal}. The dashed parts of the curve signal a negative entropy.}
    \label{fig:prova}
\end{figure}


\end{document}